%% file: 9406288.tex
\documentstyle[a4,12pt]{article}
\begin{document}
\def\seCtion#1{\section{#1} \setcounter{equation}{0}}
\renewcommand\theequation{
\ifnum\value{section}>0{\thesection.\arabic{equation}}\fi}
\input{psbox}
\begin{flushright}
CERN-TH.7262/94\\
LPTHE-Orsay-94/48\\
HD-THEP-94-19\\
FAMNSE-12-94\\
NSF-ITP-94-64\\
hep-ph/9406288
\end{flushright}
\newcommand{\be}{\begin{equation}}
\newcommand{\ee}{\end{equation}}
\newcommand{\bea}{\begin{eqnarray}}
\newcommand{\eea}{\end{eqnarray}}
\newcommand{\nn}{\nonumber}
\newcommand{\muh}{\hat\mu}
\newcommand{\dlr}{\stackrel{\leftrightarrow}{D} _\mu}
\newcommand{\vnew}{$V^{\rm{NEW}}$}
\newcommand{\vecp}{$\vec p$}
\newcommand{\dof}{{\rm d.o.f.}}
\newcommand{\prd}{Phys.Rev. \underline}
\newcommand{\pl}{Phys.Lett. \underline}
\newcommand{\prl}{Phys.Rev.Lett. \underline}
\newcommand{\np}{Nucl.Phys. \underline}
\newcommand{\vvp}{v_B\cdot v_D}
\newcommand{\dl}{\stackrel{\leftarrow}{D}}
\newcommand{\dr}{\stackrel{\rightarrow}{D}}
\newcommand{\mev}{{\rm MeV}}
\newcommand{\gev}{{\rm GeV}}
\newcommand{\calp}{{\cal P}}
\newcommand{\pinc}{\vec p \hskip 0.3em ^{inc}}
\newcommand{\pout}{\vec p \hskip 0.3em ^{out}}
\newcommand{\ptr}{\vec p \hskip 0.3em ^{tr}}
\newcommand{\pbr}{\vec p \hskip 0.3em ^{br}}
\newcommand{\no}{\noindent}
\newcommand{\ra}{\rightarrow}
\newcommand{\intsumpm} {\sum_{n^\pm}\hskip -14 pt\int\hskip 7 pt}
\newcommand{\intsump} {\sum_{n^+}\hskip -14 pt\int\hskip 7 pt}
\newcommand{\intsumm} {\sum_{n^-}\hskip -14 pt\int\hskip 7 pt}
\def\jor#1WAS#2{{%
\vrule width 1ex\raise-2pt\rlap{\vrule height0.4pt width2cm depth0pt}%
\bf#1}}
%
\def\dsl#1{\mathchoice
 {\dslaux\displaystyle{#1}} {\dslaux\textstyle{#1}} {\dslaux\scriptstyle{#1}}
 {\dslaux\scriptscriptstyle{#1}} }
\def\dslaux#1#2{\setbox0=\hbox{$#1{#2}$}
 \rlap{\hbox to \wd0{\hss$#1/$\hss}}\box0}
%
\def\frameit#1#2#3#4{\hbox{\vrule width#1\vbox{%
  \hrule height#1\vskip#3\hbox{\hskip#2\vbox{#4}\hskip#2}%
        \vskip#3\hrule height#1}\vrule width#1}}%
\let\eps\epsilon

\let\slash=\dsl

\pagestyle{empty}

\centerline{\LARGE{\bf{Standard Model CP-violation and  }}}
\vskip 1 cm
\centerline{\LARGE{\bf{Baryon asymmetry}}}
\vskip 1 cm
\centerline{\LARGE{\bf{Part I: Zero Temperature}}}

\vskip 1.5cm
\centerline{\bf{M.B. Gavela,$^a$, M. Lozano$^b$, J. Orloff$^c$,
O.P\`ene$^d$}}
\centerline{$^a$ CERN, TH Division, CH-1211, Geneva 23, Switzerland}
\centerline{$^b$ Dpto. de F\'isica At\'omica, Molecular y Nuclear, Sevilla,
Spain,\footnote{Work partially supported by Spanish CICYT, project PB
92-0663.}}
\centerline{$^c$ Institut f\"ur Theoretische Physik, Univ.
Heidelberg}
\centerline{$^d$ LPTHE, F 91405 Orsay, France,\footnote {Laboratoire
associ\'e au Centre National de la Recherche Scientifique.}}

\date{}
\begin{abstract}

We consider quantum effects in a world with two coexisting symmetry phases,
unbroken and spontaneously broken, as a result of a first order phase
transition. The discrete symmetries of the problem are discussed in
general. We compute the exact two-point Green function for a free fermion,
when a thin wall separates the two phases. The Dirac propagator displays both
massive and massless poles, and new CP-even phases resulting from the fermion
reflection on the wall. We discuss the possible quark-antiquark CP asymmetries
produced in the Standard Model(SM) for the academic $T=0$ case. General
arguments indicate that an effect first appears at order $\alpha_W$ in the
reflection amplitude, as the wall acts as a source of momentum and the
 on-shell one-loop self-energy cannot be
renormalized away. The asymmetries stem from the interference of the SM CP-odd
couplings and the CP-even phases in the propagator. We perform a toy
computation that indicates the type of GIM cancellations of the problem. The
behaviour can be expressed in terms of two unitarity triangles.

\end{abstract}

\date{}

\vfill

\newpage
\pagestyle{plain}
\seCtion{Introduction}

In ref. \cite{letter} we presented recently a summary of our ideas and results
on Standard Model (SM)
baryogenesis, in the presence of a first order phase transition. The aim of the
present work is to
describe in detail our zero temperature analysis. The finite temperature $T$
scenario is treated in the acompanying paper, ref. \cite{nousT}.

The baryon number to entropy ratio in the observed part of the universe is
estimated to be $n_B / s \sim (4-6) 10^{-11}$\cite{exp}. In 1967
A.D. Sakharov\cite{sak} established the three building blocks required from
any candidate theory of baryogenesis:
\begin{description}
\item[(a)] Baryon number violation,
\item[(b)] C and CP violation,
\item[(c)] Departure from thermal equilibrium.
\end{description}

The Standard Model (SM) contains (a)\cite{spha} and (b)\cite{KM}, while (c)
could also be large enough \cite{trans0}\cite{trans}, if a first order $SU(2)
\times U(1)$ phase transition took place in the evolution of the universe
\cite{trans2}.  We will not enter the discussion on the latter: it will be
assumed that a first order phase transition did take place, as strong as
wished by the reader.  An optimal sphaleron rate can be assumed as well. Our
aim is to argue, on a quantitative estimation of the electroweak C and CP
effects exclusively, that the current SM scenario is unable to explain the
above mentioned baryon number to entropy ratio.

Intuitive CP arguments lead to an asymmetry many orders of magnitude below
observation \cite{Jarlskog}\cite{shapo1}.  Assume a total flux of baryonic
current, where all quark flavours are equally weighted. An hypothetical
CP-violating contribution in the SM with three generations\cite{KM} should be
proportional to
\bea
{s_1}^2 s_2 s_3 c_1 c_2 c_3 s_\delta \,({m_t}^2 - {m_c}^2) ({m_t}^2 - {m_u}^2)
({m_c}^2 - {m_u}^2)&\nn\\  \times({m_b}^2 - {m_s}^2) ({m_b}^2 - {m_d}^2)
({m_s}^2 - {m_d}^2),&
 \label{cp}
\eea
times some power of the electroweak coupling constant, to be determined. In
eq. (\ref{cp}) the mixing angles and phase are the original Kobayashi-Maskawa
ones\cite{KM}. In order to obtain a dimensionless result, the expression in
(\ref{cp}) has to be divided by the natural mass scale of the problem $M_W$
or, at finite temperature, $T$, at the $12^{th}$ power. It
follows\cite{shapo1} that the resulting asymmetry is negligible, $n_B / s
\sim 10^{-20}$ or smaller if other reduction factors are considered. When the
physical boundary conditions are such that some specific flavour is singled out
by nature (or experiment as, for instance, in the $K_0 - \overline{K}_0$
system, electric dipole moment of the neutron, etc.), some of the fermionic
mass differences in (\ref{cp}) are no more compelling, and a stronger effect
is possible. Nevertheless it seems difficult to compensate the $10$ orders of
magnitude difference in the above reasoning. Such could be the case if the
desired observable came about as a ratio of (\ref{cp}) with respect to another
process which contained by itself some fermionic suppression factors: an
example is the CP-violating $\epsilon$ parameter in $K$ decays, where the
CP-violating amplitude is compared to a CP-conserving denominator, $\sim$Re
($K_0 - \overline{K}_0$), which is a GIM  suppressed object. This
is not what happens with the baryon asymmetry of the universe where, as shown
below, although nature may single out some specific flavours, there is nothing
to divide with (other than the overall normalization to the total incoming
flux, which bears no suppression factor). In this respect, the process has
some analogy with the electric dipole moment of the neutron where the
experiment preselects the up and down flavours, and of course the results are
normalized to the total incoming flux of neutrons.

On the light of the above considerations, the reader may be astonished that we
give the problem further thought. Our motivation is that the study of quantum
effects in the presence of a first order phase transition is rather new and
delicate, and traditional intuition may fail. A SM explanation would be a very
economical solution to the baryon asymmetry puzzle.  To discard this
possibility just on CP basis suggests that new physics is responsible for it,
without any need to settle whether a first order phase transition is possible
at all. Furthermore, the detailed solution in the SM gives the ``know how" for
addressing the issue in any theory beyond the standard one where CP violation
is first present at the one-loop level. On top of the above, the authors of
ref.\cite{shapo} have recently studied the issue in more detail and claim
that, at finite temperature, the SM is close to produce enough CP violation
as to explain the observed $n_B / s$ ratio.
\begin{figure}[hbt]
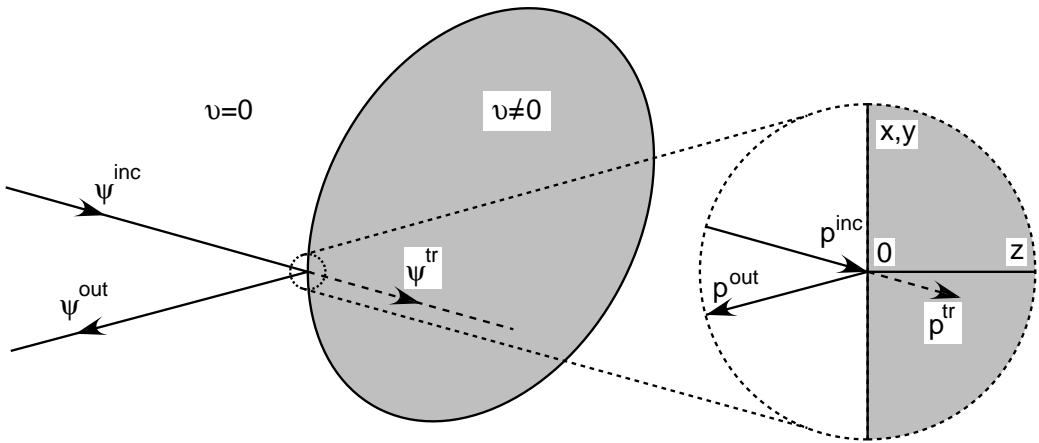
   
    \begin{center}
      \mbox{\psboxto(0.85\hsize;0 pt){bubble.ai}}
    \end{center}
    \caption{\it Quarks scattering off a true vacuum bubble. Some notations
      in the text are summarised in the ``zoomed-in" view.}
    \protect\label{wall}
\end{figure}

A first order phase transition can be described in terms of bubbles of ``true"
vacuum (with an inner vacuum expectation value of the Higgs field $v \ne 0$)
appearing and expanding in the preexisting ``false" vacuum (with $v=0$
throughout).  We can ``zoom" into the vicinity of one of the bubbles, see Fig.
\ref{wall}. There,
the curvature of its wall can be neglected and the world is divided into two
zones: on the left hand side, say, $v=0$; on the right $v\ne 0$ and masses
appear. The actual bubble expands from the broken phase ($v\ne 0$) towards the
unbroken one ($ v =0$). We work in the wall rest frame in which the plasma
flows in the opposite direction. Consider thus a baryonic flux hitting the
wall from the unbroken phase. Far enough to the left no significant
CP-violating effect is possible as all fermions are massless. In consequence,
the heart of the problem lies in the reflection and transmission properties of
quarks bumping on the bubble wall. CP violation distinguishes particles from
antiparticles and it is $a\,\, priori$ possible to obtain a CP asymmetry on
the reflected baryonic current, $\Delta_{CP}$.
The induced baryon asymmetry is at most $n_B/s\,\sim\,10^{-2}\,\Delta_{CP}$,
in a very optimistic estimation of the non-CP ingredients
\cite{linde2}\cite{shapo}.

The symmetries of the problem are analyzed in detail for a generic bubble. The
analytical results correspond to the thin wall scenario. The latter provides
an adequate physical description for typical momentum of the incoming
particles $\vert \vec p \vert$ smaller than the inverse wall thickness $l$,
i.e., $\vert \vec p \vert \ll 1/l$. For higher momenta the cutoff effects
would show up\cite{funakubo}, but it is reasonable to believe that the
accuracy of the thin wall approximation produces an upper bound for the CP
asymmetry.

The precise questions to answer in the above framework are: 1) the nature of
the physical process in terms of particles or quasi-particles responsible for
CP violation, 2) the order in the electroweak coupling constant, $\alpha_W$,
at which an effect first appears, 3) the dependence on the quark masses and
the nature of the GIM cancellations involved.

We shall consider the problem in two steps: $T=0$ in the present paper, and
finite temperature case in the subsequent one \cite{nousT}. The cosmological
first order phase transition is a temperature effect. In order to disentangle
the physical implications of the presence of a wall, with the consequent
breaking of translation invariance, from the pure thermal effects, we consider
 here an hypothetical world at zero temperature but with two phases of
spontaneous symmetry breaking. We will see that this academic model
is illuminating.

 Intuition
indicates that an existing CP violating effect already present at zero
temperature will diminish when the system is heated, because the effective
v.e.v. of the Higgs field decreases and in consequence the fermion masses do
as well (only the Yukawa couplings already present at $T=0$ remain
unchanged). The expected decrease of CP asymmetries for increasing $T$ follows,
then, from the well known fact that the Kobayashi Maskawa CP violating effects
of the standard model disappear when at least two fermion masses of the same
charge vanish or become degenerate. This intuition can be misleading only if a
new physical effect, absent at $T=0$ and relevant for the problem, appears at
finite temperature.  Treating first the $T=0$ case allows a clean analysis of
the novel aspects of physics in a world with two phases of spontaneous
symmetry breaking.  The quantum mechanical problem is well defined and in
particular the definition of particles, fields, in and out states, etc. is
transparent.

At $T=0$ the building blocks of the CP violating effect are threefold. First,
the necessary CP-odd couplings of the Cabibbo-Kobayashi-Maskawa (CKM) matrix
are at work. Second, there exist CP-even phases, equal for particles and
antiparticles, which interfere with the pure CP-odd ones to make them
observable. These are the reflection coefficients of a given particle hitting
the wall from the unbroken phase. They are complex when the particle energy is
smaller than its mass. Third, as argued below, the one loop self-energy of a
particle in the presence of the wall cannot be completely renormalized away
and results in physical transitions. The effect, thus, is present at order
$\alpha_W$ in amplitude. Such an effect is absent in the usual world with just
one phase of spontaneous symmetry breaking, where the self-energy is
renormalized away for an on-shell particle. The difference is easy to
understand: the wall acts as an external source of momentum in the one-loop
process.

The truly and essential non-perturbative effect is the wall itself. The
propagation of any particle of the SM spectrum should be exactly solved in its
presence. And this we do for a free fermion, leading to a new Feynman
propagator which replaces and generalizes the usual one. With this exact,
non-perturbative tool, perturbation theory is then appropriate in the gauge and
Yukawa couplings of fermions to bosons, and the one loop computations can be
performed.

 Strictly speaking, the gauge boson and Higgs particles propagators in the
presence of the wall are required, and it is possible to compute them with a
similar procedure\cite{petits}. Furthermore, at $T=0$ the field theoretical
problem is ill-defined in the unbroken phase where the fermions are massless,
due to infrared divergences. Some cutoff is needed, although no dependence on
it should remain when the cancellations between different diagrams contributing
to a CP-violation effect are considered.  By the time being, we work in a
simplified case in which the one-loop effect itself is computed in one of the
two phases. The final result describes a transition between any two flavours of
the same charge, resulting in a CP violating baryonic flow for a given initial
chirality, of order ${\alpha_W}^2$ in the total rate. The GIM cancellations
appear as powers of $ {m_f}^6/{M_W}^6 $ times logarithmic corrections, for the
internal fermionic masses in loops. The dependence in the external masses is
not trivial either, and its GIM implications are discussed in detail. It
follows numerically that the CP factor is orders of magnitude smaller than the
CP factor needed for baryon number generation. This $T=0$ model is academic,
and we develope it to sharpen our tools for the physical finite temperature
case\cite{nousT}.

The scope of the present paper goes beyond the particular issue of baryon
number generation in the SM. For instance, the above mentioned exact fermion
propagator in the presence of a wall, which we derive, should be useful in
other scenarios when a first order phase transition is present. The
understanding at the quantum level of physical processes in the presence of
phase transitions still is in its infancy, and we give it a
modest try. CP violation in the SM is just an example of an effect which would
disappear in a classical or semiclassical statistical physics approach.

\seCtion{Notations and symmetries}
\label{secnot}
In this section we settle the formalism and express some general
results. Consider several quark flavours in a gauge theory with a vacuum
presenting a ``wall'' structure, parallel to the $x-y$ plane. More precisely,
assume that the Higgs field (or fields if there are several Higgses) has a
vanishing v.e.v. for $z\rightarrow -\infty$ and some constant v.e.v. for
$z\rightarrow \infty$. In between there is a ``wall'' with arbitrary width and
profile. The interactions of quarks with the gauge and scalar bosons are given
by some gauge invariant Lagrangian. As previously stated, the quark-boson
interactions will be treated in perturbation, while the non-perturbative
effect of the wall on the fermion propagation is solved exactly.

As a first step let us consider the non-interacting case and treat in first
quantization formalism the free Dirac Hamiltonian, in the presence of the
wall.  The second quantized fields in the ``interaction representation'' are
built subsequently.

The quarks/antiquarks hitting the wall from the unbroken phase are reflected
or transmitted. The question is whether there exists a CP asymmetry that
produces a different  reflection probability for  quarks and
antiquarks\footnote{Note that current conservation relates transmission
to reflection, so that an
asymmetry in the reflection probability automatically implies an asymmetry in
the transmission probability.}.

In the unperturbed case, we will diagonalize the free Hamiltonian. The
eigenstates  provide then the reflection and transmission coefficients. The
fields and the quark propagators are made out of these
eigenstates. Furthermore, they will be the building blocks of the incoming
(from the unbroken phase) and outgoing (back to the unbroken phase) wave
packets when higher orders are considered later on.

The symmetries of the problem and the conserved or partially conserved quantum
numbers are relevant to classify the eigenstates. It should be obvious that
discrete symmetries are of particular relevance. In the second subsection we
shall derive in the exact theory the consequences of the only unbroken
discrete symmetry, CPT, and study in that context the effect of CP
violation. Note that the consideration of a flux of particles approaching the
wall with a non-zero average velocity (which will correspond later on to the
physical process) breaks the CPT symmetry in a global sense, but the
microscopic relations for two-particle transition amplitudes derived below
will still hold.

\subsection{Free quarks in the presence of a wall.}
\label{secfree}

Let us consider a static ``wall" parallel to the $x-y$ plane. The Higgs vacuum
 expectation value depends only on the $z$ coordinate, resulting in hermitian
 mass matrices $m(z)$ and $m_5(z)$ for the quarks.  The time independent Dirac
 Hamiltonian is then

\be H=
\vec \alpha \cdot\vec p +\beta m(z)+i\beta\gamma_5 m_5(z)\,\,, \label{dirac}
\ee
where $\alpha_i=\gamma_0\gamma_i$ and $\beta=\gamma_0$.

For the SM, in the basis where the mass matrices are diagonal, the study
simplifies to the one flavour case and $m(z)$, $m_5(z)$ become
real. Furthermore, it is always possible to rotate $m_5(z)$ away through a
chiral rotation. In other models, a diagonalization of both matrices is not
always possible and the matrix $m_5(z)$ may remain, for instance, in two-Higgs
models when the relative phase between the corresponding v.e.v.'s changes with
$z$ (see for example ref \cite{ckn}).

Let $E,p_x,p_y,p_z$ denote the four-momentum of the quark.  For a static wall
the particle energy, $E$, is conserved, while $p_z$ is not when the fermion
crosses the wall or bounces back, due to lack of translational invariance. The
system is symmetric under reflections with respect to the $x-y$ plane and with
respect to rotations around the $z$ axis, the latter implying conservation of
total angular momentum in the $z$ direction, $J_z$. It is also invariant under
Lorentz boosts parallel to the $x-y$ plane ($K_x$ and $K_y$ commute with
$H$). From the two preceding symmetries follows a conserved quantum number
for any wave function which is an eigenvector of $E,p_x$ and $p_y$.  Boost the
reference frame to obtain $p_x=p_y=0$. In that frame, the helicity states of
the incoming plane waves in the unbroken phase correspond to eigenstates of
$J_z=S_z$. This quantum number is more convenient to use than helicity because
it is conserved. Denote generically by $j_z$ ($j_z=\pm 1/2$) the corresponding
eigenvalue of $J_z$: it will be used when labeling the eigenstates of the
Hamiltonian.

Charge conjugation, C, is conserved when there exists a basis where both
$m(z)$and $m_5(z)$ are real matrices.

Define P' as the product of parity times a rotation of angle $\pi$ around the
$y$ axis, i.e., parity with respect to the $x-z$ plane, which will maintain
the unbroken phase on the initial side of the $z$ axis. P' is a symmetry when
$m_5(z)=0$ in some basis. CP' is conserved when a basis exists such that
$m(z)+i \gamma_5 m_5(z)$ is a real matrix. The unperturbed SM Hamiltonian is
invariant under C, P' and CP'. In general, the following quantum numbers are
conserved: $p_x, p_y, J_z, E$ and CP'T.

To be specific we have assumed that the unbroken phase is on the negative side
of the $z$ axis.  We define the eigenstates $\psi_n$ by
\be
E_n \psi_n= H \psi_n \label{eigen}
\ee
where the ``energy'' $E_n$ may be positive or negative, and $n$ is a short
hand notation for the quantum numbers that label the states. For the latter we
choose $n={p_x, p_y, j_z, E_n, f, \alpha}$.  $p_x, p_y$ and $E_n$ are
conserved real quantum numbers. They can take any positive or negative value
provided $E_n^2>p_x^2+p_y^2$. $j_z$ is a discrete conserved quantum number
defined above. $f$ is the flavour of the state in the broken region. $\alpha$
will be relevant when the dimension of the eigenspace is bigger than one, for
fixed values of the other quantum numbers. The eigenspace has dimension one in
the case of total reflection, i.e. when the energy is smaller than the height
of the wall, $\vert E_n \vert < m(+\infty)$, where it is assumed that the
basis is such that $m_5(\infty)=0$. In the opposite case, when the energy is
larger than the height of the wall, it has dimension two. $\alpha$ serves then
to label the state in any convenient basis. Examples will be given in the case
of the thin wall.  With the above tagging, two states with different labels
are linearly independent.

Given a complete set of orthonormal eigenstates of the Hamiltonian,
fields are defined in the usual way,
\be
\Psi(\vec x,t)= \sum_{n^+} b_{n^+}\psi_{n^+}(\vec x) e^{-iE_{n^+}t}+
\sum_{n^-}d^{\dagger}_{n^-}\psi_{n^-}(\vec x) e^{-iE_{n^-}t},\label{champs}
\ee
where $n^+$ ($n^-$) label the positive (negative) energy states.  $b_{n^+}$
and $d_{n^+}$ and their hermitian conjugates are annihilation and creation
operators\footnote{The annihilation operators verify $b_n|0>=d_n|0>=0$, the
state $|0>$ being the "vacuum" that contains the wall, i.e. with a Higgs
expectation value that depends on $z$.} with the usual anticommutation
relations. The fields $\Psi(\vec x,t)$ verify the Dirac equation in the
presence of the wall. The completeness relation for the eigenstates
$\psi_n(\vec x)$ forces upon them the standard Fermi-Dirac anticommutation
relations. They are the Heisenberg quark fields in the presence of the wall,
when the coupling of quarks to the gauge and scalar bosons are switched
off: they constitute the starting point of usual perturbation theory.

Strictly speaking, although we study transitions between two given fermions,
the propagation of bosons in the presence of the wall is needed. The reason is
that a CP-violating effect in the SM is at least a one-loop process. The
bosonic propagators in the presence of the wall are thus pertinent and viable
in analogy to the fermion treatment we are developing here\cite{petits}. For
technical simplicity, we leave such a task for a forthcoming publication, as
the subsequent modifications should not change neither the order in the
electroweak coupling constant at which the effect first shows up, nor the type
of GIM cancellations.

\begin{table}
\centering
\begin{tabular} {|c|c|c|c|c|}
\hline
 symmetry  &wave function & field & momentum& helicity \\
\hline
Identity   & $ \psi_n(\vec x)$&$
\Psi(\vec x,t) $&$ p$ & h   \\
$P'$   & $ i\sigma_y \gamma_0 \psi_n(\tilde x)$&$ i\sigma_y \gamma_0
\Psi(\tilde
x,t) $&$\tilde p$ & -h   \\
$C$   & $ i \gamma_2 \psi_n^*(\vec x)$&$ i \gamma_2 \Psi^*(\vec
x,t) $&$ p$ & h   \\
$CP'$   & $  \gamma_5 \psi_n^*(\tilde x)$&$ \gamma_5 \Psi^*(\tilde
x,t) $&$\tilde p$ & -h   \\
 $T$   & $ -\sigma_y \psi_n^*(\vec x)$&$ -\sigma_y \Psi(\vec
x,-t) $&$ - p$ & h   \\
$TCP'$   & $  -\sigma_y\gamma_5 \psi_n(\tilde x)$&$ -\sigma_y\gamma_5
\Psi^*(\tilde
x,-t) $&$-\tilde p$ & -h   \\
\hline
\end{tabular}
\caption{\it{Transformation laws for the wave functions and the fields under
the discrete symmetries of the wall, with $\tilde p= (p_x, -p_y, p_z)$. In the
case of the symmetries C, CP' and CP'T the transformed wave function
corresponds to a negative energy state. However the transformed momenta and
helicities we quote are those of the anti-particles, i.e. opposite to those
for negative energy states.}}
\label{tab:symm}
\end{table}

Table \ref{tab:symm} shows the transformation laws for the discrete symmetries
of the system.

\subsection{Symmetries of the exact theories.}
\label{secexact}

Consider now the exact theory with the quark-boson interactions
incorporated. The states in eq. (\ref{eigen}) are no more eigenstates of the
total Hamiltonian. The fields defined in eq. (\ref{champs}) are not Heisenberg
fields of the full theory, but they can be used as the quark fields in the
``interaction representation''. The discrete symmetries apply to the fields
exactly as stated in Table \ref{tab:symm}.  In general all these symmetries
are broken {\it except} TCP'.

Consider, for definiteness, the unbroken phase. {}From Table \ref{tab:symm} the
TCP'-transformed of a state with flavour $f$, momentum $\vec p$ and helicity
$h$ is given by
\be
|f, \vec p, h>^{TCP'} = |\bar f, -\tilde p, -h>,
\label{TCP1}
\ee
where $\bar f$ corresponds to the antiquark with the same flavour.
TCP' is an antiunitary operation, and  invariance under it implies that

\bea  ^{TCP'}<b |e^{-iH(t'_1-t'_2)}|a>^{TCP'} &=&
\bigl ( <b |e^{-iH(t_1-t_2)}|a>\bigr )^* \nn \\
&=& <a |e^{-iH(t'_1-t'_2)}|b>,
\label{TCP2}\eea

where $a,b$ are two given quark states and $t'_1=-t_1$ and $t'_2=-t_2$.  The
relation $e^{-iH(t'_1-t'_2)}=
(e^{-iH(t_1-t_2)})^\dagger$ has been used. Equations (\ref{TCP1}) and
(\ref{TCP2})
imply, for instance for a top to charm transition, that
\bea
<c,(p_x,p_y,-p_z),R|e^{-iH(t_1-t_2)}|t,\vec p,L>=&\nn\\
<\bar{t},(-p_x,p_y,-p_z),R&|e^{-iH(t_1-t_2)}|\bar{c},(-p_x,p_y,p_z),L>,
\eea
and the analogous formula for all flavours and helicities.

Letting $t_1 \rightarrow \infty$ and $t_2 \rightarrow -\infty$, two important
consequences follow from this equation:
\begin{itemize}
\item {i)} All CP' asymmetries obtained by summing over all flavours,
helicities and incoming momenta will necessarily vanish\cite{ckn}:
\be
\sum_{f,f', h, p^{inc}}
  \vert A^2(   f,h,p^{inc}\rightarrow      f',-h,p^{out})\vert^2 -
  \vert A(\bar f,h,p^{inc}\rightarrow \bar f',-h,p^{out})\vert^2 = 0,
\label{asym1}
\ee
where $A(f,h,p^{inc}\rightarrow f',-h,p^{out})$ represents the amplitude for
an incoming particle $f$ with helicity $h$ and moemntum $ p^{inc}$ to be
reflected into an outgoing particle
$f'$ with helicity $ -h$ and momentum $p^{out}=(p_x^{inc},p_y^{inc},
-p_z^{inc},E)$. This strong result does not  kill, however, the  scenario
for baryogenesis during the electroweak phase transition, since the latter
is based on the role of sphalerons, which is totally different for left-handed
and right-handed quarks. In the following we will consider CP' asymmetries
summed over flavours and momenta but only for a given initial helicity.

\item {ii)} A straightforward consequence of equation \ref{asym1} is
\bea
\sum_{f,f', p^{inc}} &\vert A(f,L,p^{inc}\rightarrow f',R,p^{out}) \vert^2 -
\vert A(\bar f,R,p^{inc}\rightarrow \bar f',L,p^{out}) \vert^2 = \nn \\
\sum_{f,f', p^{inc}} &\vert A(f,L,p^{inc}\rightarrow f',R,p^{out}) \vert^2 -
\vert A( f,R,p^{inc}\rightarrow  f',L,p^{out}) \vert^2,
\label{asym2}\eea
which is a useful simplification, as it allows to express a CP' asymmetry
without any use of anti-quarks.
\end{itemize}

As we are interested in the breaking of CP' symmetry, it is useful to
express the consequences of an hypothetical theory invariant under this
transformation:
\bea
<c,(p_x,p_y,-p_z,R\vert e^{-iH((t_1-t_2)}\vert t,\vec p,L>=&\nn\\
 <\bar c,(p_x,-p_y,-p_z,L\vert &e^{-iH((t_1-t_2)}\vert \bar
t,(p_x,-p_y,p_z),R>,
\label{CP}
\eea
and the analogous formulae for all flavours and helicities.

\seCtion{The thin wall.}
\label{sectw}
{}From now on we consider the SM in the presence of a thin wall, which allows
to
perform easier analytic calculations. The unperturbed Hamiltonian is, for each
flavour,
\be
H= \vec \alpha \cdot\vec p +\beta m \theta(z). \label{thin}
\ee

\subsection{Eigenstates}
\label{seceig}
Consider a positive energy fermion with given $p_x,p_y,E$.\footnote{
$E>\sqrt{p_x^2+p_y^2}$.}

On the left of the wall ($z < 0$), an incoming particle has a
$z$-momentum $p_z=\sqrt{E^2-p_x^2-p_y^2}>0$, and helicity $h$. Define
\be
p^{inc}=(p_x, p_y, p_z, E).\label{minc}
\ee
Upon hitting the wall, a reflected particle has $z$-momentum $-p_z$ and
\be
p^{out}=(p_x, p_y, -p_z, E),\label{mout}
\ee
while a transmitted one has $z$-momentum $p'_z$ given by
\bea p'_z = \sqrt{p_z^2-m^2} \, \, & \hbox{if} \, \, \, \, p_z^2 > m^2 \nn \\
     p'_z = i\sqrt{m^2-p_z^2} \, \, & \hbox{if} \, \,\, \, p_z^2 < m^2.
\label{pprime}
\eea

With these definitions, $e^{ip'_z\cdot z}$ is a falling exponential in the
case of total reflection ($p_z^2 < m^2$). We thus define the transmitted
4-momentum as
\be
p^{tr}=(p_x, p_y, p'_z, E).\label{mtr}
\ee
The above physical situation is described by the following eigenstate of the
hamiltonian, which will be denoted ``incoming" eigenstate,
\bea
\psi_{n^+}^{inc}(\vec x)=& \bigg (u_h({\pinc})\,e^{i{\pinc}\cdot \vec x}
+ R\, u_h({\pinc})\,e^{i{\pout}\cdot \vec x}\bigg )\,\,\theta(-z)\nn \\&
+ (1+R)\, u_h({\pinc})\,e^{i{\ptr}\cdot \vec x}\,\,\theta(z),\,\,
\quad h=j_z\,\,.
\label{psin}
\eea
It is directly related to an incoming wave packet. $u_h({\pinc})$ is a
solution of the  Dirac equation in the unbroken phase,
\be
\slash p^{inc} u_h({\pinc})=0, \label{dirinc}
\ee
and the reflection matrix $R$ is given by
\be
R=\frac {m\gamma_z} {p_z+p'_z}.\label{R}
\ee

The Dirac structure of $R$ reflects the opposite chirality of the incoming and
reflected state. Note as well that $R$ is complex in the case of total
reflection.  Its imaginary phase does not change from particles to
antiparticles, and constitutes an example of the CP-even phases which will
later
on interfere with the CP-odd ones to produce observable effects.

It follows from the definition of $R$ that $(1+R)u_h^{inc}({\pinc})$
is a solution of the Dirac equation in the broken phase,
\be
(\slash p^{tr}-m)\,(1+R)\,u_h({\pinc})=0\,\,,\label{dirtra}
\ee
as well as
\be
\slash p^{out}\,R\,u_h({\pinc})=0\,\,.\label{dirout}
\ee

In fact, eq. (\ref{dirtra}) was used to compute $R$.  In the case of total
reflection, $p_z^2-m^2<0$, the wave function (\ref{psin}) is the only positive
energy eigenstate.

When transmission occurs, one more eigenstate is needed to span the
eigenspace. A second linearly independent solution, to be called ``outgoing''
state (because directly related to the outgoing wave packets), is given by
\bea
\psi_{n^+}^{out}(\vec x)=& \bigg (u_h({\pout})\,e^{i{\pout}\cdot \vec x}
+  R^\dagger\, u_h({\pout})\,e^{i{\pinc}\cdot \vec x}\bigg )\,\,
 \theta(-z)\nn \\&
+ (1+ R^\dagger)\, u_h({\pout})\,e^{i{\pbr}\cdot \vec x}\,\,
\theta(z),\,\,\quad h=-j_z\,,
\label{psout}
\eea
where
\be
p^{br}=(p_x, p_y, -(p'_z)^*, E).\label{mbr}
\ee

This wave function $\psi^{out}$ is however not orthogonal to $\psi^{inc}$.  An
orthogonal eigenstate is provided by the solution coming from the broken
phase:
\bea
\psi_{n^+}^{br}(\vec x)=& \sqrt {p_z/p'_z} \bigg [ \bigg (u_s({\pbr})
  e^{i{\pbr}\cdot \vec x}
+ J u_s({\pbr})e^{i{\ptr}\cdot \vec x}\bigg )\theta(z)\nn \\&
+ (1+J) u_s( {\pbr})e^{i {\pout}\cdot \vec x}\theta(-z) \bigg ],
\label{psbrok}\eea
where $J$ is the reflection matrix for a particle bouncing on the wall from
the broken phase, given by
\be
1+J = \frac{p'_z}{p_z}(1+R),\label{J}
\ee
and $s$ is a spin index dependent on $j_z$, such that $(1+J) u_s(
{\pbr})=u_h(\pout)$ is a massless spinor with helicity $h=-j_z$, and
$u_s(\pbr)$ satisfies the Dirac equation in the broken phase
\be (\slash p^{br}-m)\,u_s({\pbr})=0\,\,.\label{diria}\ee

For later reference, we will name the solution in eq. (\ref{psbrok}) the
``broken incoming'' wave function.

The negative energy solutions, $\psi_{-n}$, describe the propagation of
antiparticles. In terms of the quantum numbers for positive energy
antifermions,
the corresponding wave functions are
\bea
\psi_{n^-}^{inc}(\vec x)=& \bigg (v_h({\pinc})\,e^{i{\pinc}\cdot \vec x}
+\overline{R}\, v_h({\pinc})\,e^{i{\pout}\cdot \vec x}\bigg )\,\,\theta(-z)
  \nn \\
&+ (1+\overline{R})\, v_h({\pinc})\,e^{i{\ptr}\cdot \vec
  x}\,\,\theta(z),\,\,\quad h=j_z\,,
\label{psina}\eea
\bea \psi_{n^-}^{out}(\vec x)=& \bigg (v_h({\pout})\,e^{i{\pout}\cdot \vec x}
+ \overline{R}^\dagger\, v_h({\pout})\,e^{i{\pinc}\cdot \vec x}\bigg)\,\,
  \theta(-z)\nn \\
& + (1+ \overline{R}^\dagger)\,
v_h({\pout})\,e^{i{\pbr}\cdot \vec x}\,\,\theta(z),\,\,\quad h=-j_z,
\label{psouta}\eea
and
\bea
\psi_{n^-}^{br}(\vec x)=
& \sqrt {p_z/p'_z}\bigg [ \bigg (v_s({\pbr})e^{i{\pbr}\cdot \vec x}
+ \overline{J} v_s({\pbr})e^{i{\ptr}\cdot \vec x}\bigg )\theta(z)\nn \\
&+ (1+\overline{J}) v_s( {\pbr})e^{i {\pout}\cdot \vec x}\theta(z) \bigg ],
\label{psbroka}
\eea
where $\overline{R}= - R$ and $(1+\overline{J})=
p'_z/p_z(1+\overline{R})$. $v_h({\pinc})$ is a negative energy Dirac spinor
verifying the equation
\be
\slash p^{inc} v_h({\pinc})=0.\label{dirincm}
\ee
with the helicity $h$ defined as $\vec\sigma\cdot \hat p\,v_h(\vec
p)=-h\,v_h(\vec p)$. The spin $s$ in eq. (\ref {psbroka}) is defined as for
eq. (\ref {psbrok}), in such a way that $(1+\overline{J}) v_s(
{\pbr})=v_h(\pout)$ with helicity $h=-j_z$.  It follows from the definition of
$\overline{R}$ that $(1+\overline{R})v_h({\pinc})$ is a solution of the Dirac
equation in the broken phase:
\be
(\slash p^{tr}+m)\,(1+\overline{R})\,v_h({\pinc})=0,\label{dirtram}
\ee
as well as
\be
\slash p^{out}\,\overline{R}\,v_h({\pinc})=0,\label{diroutm}
\ee
and
\be
(\slash p^{br}+m)\,v_s({\pbr})=0.\label{dirbro}
\ee

It is convenient to consider the Fourier transformed wave functions
\be
\widetilde\psi(q)=\int {dz\over 2\pi} e^{-iqz} \psi(z).
\ee
{}From now on it will be assumed that the boost leading to $p_x=p_y=0$ has been
performed.  We will only write here the $z$ dependence, since the $x-y$ part is
trivial ($\delta(q_x)\delta(q_y)$). The results are
\be
\widetilde\psi_{n^+}^{inc}(q_z)=
 \frac{1}{2i\pi}\bigg\{\frac{1}{p_z-(q_z+i\epsilon)}
-\frac{R}{p_z+q_z+i\epsilon}-\frac{1+R}{p'_z-(q_z-i\epsilon)}\bigg\}u_h(\pinc),
\label{fpsin}
\ee
\be
\widetilde\psi_{n^+}^{out}(q_z)=
  \frac{1}{2i\pi}\bigg\{\frac{1}{-p_z-(q_z+i\epsilon)}
-\frac{R^\dagger}{-p_z+q_z+i\epsilon}
-\frac{1+R^\dagger}{-(p'_z)^*-(q_z-i\epsilon)}\bigg\}u_h(\pout),
\label{fpsout}
\ee
\be
 \widetilde\psi_{n^+}^{br}(q_z)= \frac{1}{2i\pi}
 \sqrt{p_z/p'_z}\bigg\{\frac{1}{-p'_z-(q_z-i\epsilon)}
-\frac{J}{p'_z-(q_z-i\epsilon)}
-\frac{1+J}{-p_z-(q_z+i\epsilon)}\bigg\}u_s(\pbr),
\label{fpsrok}
\ee
\be
\widetilde\psi_{n^-}^{inc}(q_z)=
\frac{1}{2i\pi}\bigg\{\frac{1}{p_z-(q_z+i\epsilon)}
-\frac{\overline{R}}{p_z+q_z+i\epsilon}
-\frac{1+\overline{R}}{p'_z-(q_z-i\epsilon)}\bigg\}v_h(\pinc),
\label{fpsina}
\ee
\be
 \widetilde\psi_{n^-}^{out}(q_z)=
 \frac{1}{2i\pi}\bigg\{\frac{1}{-p_z-(q_z+i\epsilon)}
-\frac{\overline{R}^\dagger}{-p_z+q_z+i\epsilon}
-\frac{1+\overline{R}^\dagger}{-(p'_z)^*-(q_z-i\epsilon)}\bigg\}v_h(\pout),
\label{fpsouta}
\ee
\be
 \widetilde\psi_{n^-}^{br}(q_z)= \frac{1}{2i\pi} \sqrt{p_z/p'_z}
\bigg\{\frac{1}{-p'_z-(q_z-i\epsilon)}
-\frac{\overline{J}}{p'_z-(q_z-i\epsilon)}
-\frac{1+\overline{J}}{-p_z-(q_z+i\epsilon)}\bigg\}v_s(\pbr),
\label{fpsroka}
\ee
with $h=+j_z$ in $\widetilde\psi^{inc}_{n^\pm}$ and $h=-j_z$ in
$\widetilde\psi^{out}_{n^\pm}$.

\subsection{Wave packets}
\label{secwp}
The physical process under study is described by a wave packet coming from the
unbroken phase, bouncing on the wall, and generating reflected and transmitted
wave packets.  The transmission probability follows from the reflection one.
It is thus sufficient to concentrate on reflection properties.

Each wave packet has fixed flavour and helicity $h$ (equivalent in the unbroken
phase to $j_z$, $-j_z$, respectively for the incoming, outgoing wave
packets). Its energy and momenta are clustered around average
values. Somewhere in the process the electroweak interactions will act as a
perturbation, and induce flavour changes through loops, as will be discussed in
Section \ref{secloop}.

The $x$ and $y$ components of the wave packet are not modified by the wall,
and will be ignored in what follows. Consider the following incoming wave
packet, approaching the wall from the unbroken phase,
\bea
P(\pinc,z,t)&=&N\int dk_z \,e^{-(k_z-p_z^{inc})^2\frac{d^2}{2}}
\,e^{ik_z(z+Z-t-\cal{T})}\,u_h(\vec k)\nn \\
&\sim&e^{ip_z^{inc}(z+Z-t-\cal{T})}\,e^{-\frac{(z+Z-t-\cal{T})^2}{2d^2}}
\,u_h(\pinc),
\label{wp}
\eea
where $N$ is a normalization constant, $d$ denotes the spatial extension of
the wave packet, which is located at time $\sim{-\cal{T}}$ ($\cal{T}>0$) around
position
$\sim -Z$ ($Z>0$), assuming $Z/d>>1$, $T/d>>1$. $d$ is chosen so that
$p_z^{inc}d>>1$.  Terms exponentially suppressed by $\exp{[-(p_z^{inc}d)^2]}$
have been neglected in the second line of eq. (\ref{wp}).

At $t \sim -\cal{T}$, the wave packet is an almost monochromatic wave located
far
from the wall in the unbroken phase, with group velocity pointing towards the
wall. To study its time evolution when approaching the wall, an expansion on
the eigenstates defined in subsection \ref{seceig} is convenient. Up to
exponentially suppressed terms, the result is
\be
P(\pinc,z,t)= N\int dk_z \,e^{-(k_z-p_z^{inc})^2\frac{d^2}{2}}\,
\psi_{n^+}^{inc}(z)\,e^{-ik_z(t+\cal{T})}.\label{wp2}
\ee

In other words, the incoming wave packet totally expands on the eigenstates
denoted by $\psi_{n^+}^{inc}$, thus justifying their name. Analogous
conclusions hold for the outgoing wave packet in the unbroken phase. It
follows that the physical amplitude can be expressed at order $\alpha_W$ as
\be
A(i\rightarrow f) = <\psi^{out}_n| H_{int}| \psi^{inc}_n>,
\label{wp3}
\ee
where $H_{int}$ is the interaction Hamiltonian induced by electroweak loops.

\seCtion{The quark propagator.}
\label{secprop}
The free quark propagator is defined as usual:
\be
S(x,y)=<0\vert T\big (\Psi(x) \overline \Psi(y) \big )\vert 0>.
\label{prop0}\
\ee
It is not translational invariant due to the presence of the wall, and verifies
the equation
\be
(\partial_{x_0}+iH(x)) S(x,y)=\gamma_0 \delta_4(x-y),\label{prop2}
\ee
where $H(x)$ is the Hamiltonian in eq.  (\ref{thin}). Furthermore, the time
ordered product in equation (\ref{prop0}) insures the Feynman boundary
conditions. From eq. (\ref {champs}), we get
\bea
S(x,y)\gamma_0=\big (\theta(x_0-y_0)\intsump
&\psi_{n^+}(\vec x)\psi^\dagger_{n^+}(\vec y)
  e^{-iE_{n^+}(x_0-y_0)} - \nn \\
\theta(y_0-x_0)\intsumm
&\psi_{n^-}(\vec x)\psi^\dagger_{n^-}(\vec y)
 e^{-iE_{n^-}(x_0-y_0)}\big ),
\label{prop3}\eea
where $E_{n^\pm}$ is the positive/negative energy of the eigenstate,
cf. eq. (\ref{eigen}), and where the  definition
\be
\intsumpm\equiv
\sum_{j_z,f,\alpha}\left(\frac 1{2\pi}\right)^3 \int dp_x dp_y dE_{n^\pm}
\ee
has been used.

The Fourier transformed propagator depends on two momenta, $q^i$ and $q^f$,
unlike in the translationally invariant case,
\bea
& q^f = (E,q_x,q_y,q^f_z), \nn \\ & q^i=(E,q_x,q_y,q^i_z).
\label{mom12}\eea
We can write
\bea
S(q^f,q^i)&=&\int d\xi_z d\xi'_zd(\xi'_x-\xi_x)
  d(\xi'_y-\xi_y) d(\xi'_0-\xi_0)\quad S(\xi',\xi)\times\nn\\
& &\qquad e^{-i q^f_z \xi'_z +iq^i_z
 \xi_z -iq_x(\xi'_x-\xi_x)
-iq_y(\xi'_y-\xi_y)+iE(\xi'_0-\xi_0)},
\label{four}
\eea
where translation invariance in $x, y $ and time directions has been used. From
now on we use the liberty of boosting in the $x, y$ plane to bring $q_x$ and
$q_y$ to 0. At the end of the section we will return to the general case.

In terms of the momentum-space wave functions, the above translates into
\bea
S(q^f,q^i) \gamma_0 =  \frac{-1}{i}
\intsumpm  (2\pi)^6 & \bigg[
\widetilde\psi^{inc}_{n^+}(q^f_z)
 \left( \widetilde\psi^{inc}_{n^+}(q^i_z)\right)^\dagger
 \frac {-1} {E-E_{n^+} + i \epsilon} \nn \\
&+ \widetilde\psi^{inc}_{n^-}(q^f_z)
 \left( \widetilde\psi^{inc}_{n^-}(q^i_z)\right)^\dagger
 \frac {-1} {E-E_{n^-}- i \epsilon} \nn \\
&+ \widetilde\psi^{br}_{n^+}(q^f_z)
 \left( \widetilde\psi^{br}_{n^+}(q^i_z)\right)^\dagger
 \frac {-1} {E-E_{n^+} + i \epsilon} \nn \\
&+ \widetilde\psi^{br}_{n^-}(q^f_z)
 \left( \widetilde\psi^{br}_{n^-}(q^i_z)\right)
 \frac {-1} {E -E_{n^-} - i \epsilon} \bigg].
\label{ay}\eea
Notice in the above expression that an orthonormal basis for the wave functions
was used, i.e., the one spanned by $\widetilde\psi^{inc}$
and $\widetilde\psi^{br}$, given in eqs. (\ref{fpsin}),
(\ref{fpsina}),(\ref{fpsrok}) and (\ref{fpsroka}). In particular,
orthogonality requires to restrict the energy integration on the broken phase
to
$E>m$. We have explicitly verified the completeness relation for
our system of eigenstates\footnote{ The negative energy eigenfunctions
describe the propagation of antiparticles: in the notation presently used an
initial antiparticle will have momentum $q^f$, while a final antiparticle has
momentum $q^i$.},
\bea
&&\intsump \bigg [
 \widetilde\psi^{inc}_{n^+}(q^f_z)
 \left(\widetilde\psi^{inc}_{n^+}(q^i_z)\right)^{\dagger} +
 \widetilde\psi^{br}_{n^+}(q^f_z)
 \left( \widetilde\psi^{br}_{n^+}(q^i_z)\right)^{\dagger}\bigg]\label{compl}\\
&+&\,\intsumm\bigg[
\widetilde\psi^{inc}_{n^-}(q^f_z)
\left(\widetilde\psi^{inc}_{n^-}(q^i_z)\right)^{\dagger} +
\widetilde\psi^{br}_{n^-}(q^f_z)
\left(\widetilde\psi^{br}_{n^-}(q^i_z)\right)^{\dagger} \bigg ]
\,=\frac 1{(2\pi)^3}\,\delta(q_z^f-q_z^i).\nn
\eea
The obvious consequence of eqs.  (\ref{eigen}) and (\ref{compl}) is the
Green's function equation for the propagator:
\be
-i(E-H)\, S(q^f,q^i) \gamma_0= (2\pi)^3\delta(q_z^f-q_z^i).\label{green}
\ee

The Fourier transform of
eq. (\ref{green}) is:
\be
\left(\partial_{\xi^0} +\vec\alpha\cdot\vec\nabla_{\vec\xi}+i\beta m
\theta(\xi_z)\right)S(\xi,\xi') = \gamma_0\delta^4(\xi-\xi').\label{green2}
\ee

Details on the computation of the propagator can be found in
Appendix \ref{app-proof}. Let us summarise the results in
1+1 dimensions (time and $z$-directions). An interesting intermediate step is
given by the expressions which follow. The contribution of states propagating
exclusively on the left-hand side of the wall ($v=0$, ``unbroken"
region) is
$$ S(q^f,q^i)_{left} \gamma_0 = \bigg\{
\frac {1} {q^f_z - q^i_z + i\epsilon}
 \frac {1} {(sE+i\epsilon)^2 - (q^f_z +i\epsilon)^2}
  (E+q^f_z\alpha_z)
\nn
$$
$$
-\frac{1}{(sE + i\epsilon)-(q^f_z+i\epsilon)}
 \frac {1} {sE-q^i_z + i\epsilon}
 \frac {s(1+s\alpha_z )} {2}
\nn
$$
\be
+\frac {1} {sE +q^f_z +i\epsilon}
 \frac {1} {sE-q^i_z + i\epsilon}
 \frac {(1-s\alpha_z) }{2}
 \frac {sm\gamma_0}{sE+p'_0}\bigg\}.
\label{eleft}
\ee
The contribution from states propagating exclusively on the
right-hand side of the wall (v.e.v. $\ne 0$), that is, in the ``broken" region
is
$$ S(q^f,q^i)_{right} \gamma_0 = \bigg\{
\frac {-1} {q^f_z - q^i_z - i\epsilon}
 \frac {1} {(p_z'+i\epsilon)^2 - (-q^f_z +i\epsilon)^2}
  (E+q^f_z\alpha_z+ m\gamma_0)
\nn
$$
$$
-\frac{1}{(p_z' + i\epsilon)+(q^f_z-i\epsilon)}
 \frac {1} {p_z'+q^i_z + i\epsilon}
 \frac {E} {p_z'}
 \frac12
 \left(1-\frac{p_z'}{E} \alpha_z + \frac{m}{E}\gamma_0\right)
\nn
$$
\be
-\frac {1} {p_z' -q^f_z +i\epsilon}
 \frac {1} {p_z' +q^i_z +i\epsilon}
 \frac {E} {p_z'}
 \frac12
 \left(1+\frac{p_z'}{E} \alpha_z +
 \frac{m}{E}\gamma_0\right) \frac {sm\gamma_0}{sE+p'_0}\bigg\}.
\label{eright}
\ee
Finally, the contribution from states propagating across the wall is
$$
 S(q^f,q^i)_{across}\gamma_0 = s \bigg\{
 \frac{1}{p_z'-q^f_z+ i\epsilon}
  \frac{1}{sE-q^i_z+i\epsilon}
  \left(1+ \frac {sm\gamma_0}{sE+p'_0}\right)
  \frac {(1+s\alpha_z )} {2}
\nn
$$
\be
+\frac{1}{sE+q^f_z+ i \epsilon}
  \frac{1}{p_z'+q^i_z+i \epsilon}
  \frac {(1-s\alpha_z )}{2}
  \left(1+ \frac {sm\gamma_0}{sE+p'_0}\right)
\bigg\},
\label{ecross}
\ee
where $s$ is a parameter which takes values $+1$ or $-1$ for positive or
negative energy particles, respectively. $p'_0$ is defined as follows
\be
p'_0 = +\sqrt{E^2 - m^2} + i\epsilon,
\label{ep'0}
\ee
and the factor $ s m \gamma_0 /(sE+p'_0)$ originates from the reflection
matrix $ {R}$, eq. (\ref{R}).

Eqs. (\ref{eleft})-(\ref{ecross}) show sometimes two $i\epsilon$ terms with
opposite signs present in the same denominator, for instance ${1}/({(sE +
i\epsilon)-(q^f_z+i\epsilon)})$ in the second line of eq. (\ref{eleft}). The
reason for this phenomenon is clear. The first $i\epsilon$ in our example is
there to implement the Feynman boundary conditions for the propagator in time
direction. The second $i\epsilon$ specifies in which spatial phase ($z<0$ or
$z>0$) this terms acts. This may seem ambiguous: is the pole located in the
upper or lower half plane? Both prescription are not equivalent, unless the
residue at the pole vanishes. In other words, since $1/(x\pm
i\epsilon)=P(1/x)\mp \pi\delta(x)$, the ambiguity disappears only if the factor
multiplying $\delta(x)$ vanishes at $x=0$. A close scrutiny of
eqs. (\ref{eleft})-(\ref{ecross}) shows that this is indeed the case. For
instance, the residue at $sE=q^f_z$ in the second line of eq. (\ref{eleft}) is
exactly compensated by the residue of the equally ``ambiguous'' term in the
first line of eq. (\ref{eleft}). It results that the sum of both terms is not
 ambiguous. It is possible to chose {\it in a consistent way} any sign
for the $i\epsilon$ in the first and second lines of
eq. (\ref{eleft}). Flipping from one convention to the other exchanges a term
proportional to $\delta(sE-q^f_z)$ between the first and second line of
eq. (\ref{eleft}), and the sum does not change. For later use denote ``time
$i\epsilon$'' the usual Feynman convention and ``space $i\epsilon$''
convention the other one. In the following the ``time $i\epsilon$'' convention
will be assumed unless specified otherwise.

The propagator is given by the sum of eqs. (\ref{eleft}), (\ref{eright}) and
(\ref{ecross}) above,
\be
S(q^f,q^i)\gamma_0=\bigg(S(q^f,q^i)_{left}+
S(q^f,q^i)_{right}+S(q^f,q^i)_{across}\bigg )\gamma_0.
\label{sfeo}
\ee
In Appendix \ref{app-proof}  it is explicitly shown that
eq. (\ref{sfeo}) verifies the Dirac equation.
Using the relations (\ref{identunbr}-\ref{identbr}), as well as the above
mentioned freedom to shift certain poles, it is possible to combine the
various terms in a more compact and familiar way. One can either show that

$$
S(q^f,q^i)=-\bigg\{
\frac{1}{q^f_z-q^i_z+i\epsilon} \frac{1}{\dsl q^i} -
 \frac{1}{q^f_z-q^i_z-i\epsilon} \frac{1}{\dsl q^i-m}+
\nn
$$\nobreak
$$
\frac{1}{\dsl q^f}
 \left[\frac{1+s \alpha_z}{2}\right]
 \left[1-\frac{m s \gamma_0}{E+p_z'}\right] s \gamma_0
 \frac{m}{\dsl q^i(\dsl q^i-m)} -
\nn
$$\nobreak
\be
\frac{1}{\dsl q^f-m}
 \left[1-\frac{m s \gamma_0}{E+p_z'}\right]
 \left[\frac{1- s \alpha_z}{2}\right] s \gamma_0
 \frac{m}{\dsl q^i(\dsl q^i-m)} \bigg\},
\label{prol}
\ee
which simplifies the action of the $q^i$-Dirac operator, or that

$$
S(q^f,q^i)=-\bigg\{
\frac{1}{q^f_z-q^i_z+i\epsilon} \frac{1}{\dsl q^f} -
 \frac{1}{q^f_z-q^i_z-i\epsilon} \frac{1}{\dsl q^f-m}+
\nn
$$\nobreak
$$
\frac{m}{\dsl q^f(\dsl q^f-m)}
 \left[1-\frac{m s \gamma_0}{E+p_z'}\right]
 \left[\frac{1- s \alpha_z}{2}\right] s \gamma_0
 \frac{1}{\dsl q^i} -
\nn
$$\nobreak
\be
\frac{m}{\dsl q^f(\dsl q^f-m)}
 \left[\frac{1 +s \alpha_z}{2}\right]
 \left[1-\frac{m s \gamma_0}{E+p_z'}\right]  s \gamma_0
 \frac{1}{\dsl q^i-m} \bigg\},
\label{pror}\ee
which prepares for the action of the $q^f$-Dirac operator. In fact,
eq. (\ref{prol}) is the CP' conjugate of (\ref{pror}).
An elegant expression, because more symmetric in the $q^i$ and $q^f$
dependences, can be obtained combining (\ref{prol}) and (\ref{pror}) and is
given by
$$
\frameit{0.4pt}{10pt}{0pt}{
\vbox{\multiply\hsize by 5\divide\hsize by 6
$$
S(q^f,q^i)=-1/2\bigg\{
\frac{1}{q^f_z-q^i_z+i\epsilon}
 \left(\frac{1}{\dsl q^f}+\frac{1}{\dsl q^i}\right) -
 \frac{1}{q^f_z-q^i_z-i\epsilon}
\nn
$$\nobreak
$$
 \left(\frac{1}{\dsl q^f-m}+\frac{1}{\dsl q^i-m}\right) +
\frac{1}{\dsl q^f-m}\gamma_z \frac{1}{\dsl q^i}-
 \frac{1}{\dsl q^f}\gamma_z \frac{1}{\dsl q^i-m}-
\nn
$$\nobreak
\be
\frac{m}{\dsl q^f(\dsl q^f-m)}
 \left[1-\frac{m s \gamma_0}{E+p_z'}(1-s\alpha_z)\right] s \gamma_0
 \frac{m}{\dsl q^i(\dsl q^i-m)}\bigg\}\nn
\label{propsym}\ee
}}
$$

In (\ref{prol}), (\ref{pror}) and (\ref{propsym}): all $\dsl q$'s in the
denominators are understood as regularized ``\`a la Feynman'', i.e. as $\dsl q
+ i\epsilon$, the ``time $i\epsilon$'' convention. This convention allows to
check easily that our propagator obeys Feynman boundary conditions, i.e. that
the positive frequencies ($q_0 >0$) are ``retarded'', and the negative ones
($q_0 <0$) ``advanced''.

If the ``space $i \epsilon$" convention was used instead, then the
regularisation implies to consider all massless poles ($1/\dsl q^f, 1/\dsl
q^i$) with $q^f_z\ra q^f_z + i \epsilon$ and $q^i_z\ra q^i_z - i
\epsilon$. Upon Fourier transform it corresponds to $\theta(-z_i)$ and
$\theta(-z_f)$ respectively, i.e. it locates the massless poles in the
unbroken region. Analogously, all massive poles ($1/(\dsl q^f-m_f), 1/(\dsl
q^i-m_i)$) are to be understood as $q^f_z\ra q^f_z - i \epsilon$ and $q^i_z\ra
q^i_z + i \epsilon$, locating them in the broken region. In this convention it
is easy to check that the propagator in eqs. (\ref{prol}), (\ref{pror}) and
(\ref{propsym}) verifies the Dirac equation (\ref{green2}).

For some terms in eqs. (\ref{prol}), (\ref{pror}) and (\ref{propsym}) the
``time $i\epsilon$" and the ``space $i\epsilon$'' conventions are apparently
in conflict. For instance,
\be
\frac {\dsl q^f}{q_0^2-(q_z^f+i \epsilon)^2}=
\frac {\dsl q^f}{q_0^2-(q_z^f)^2+i \epsilon}
+ 2 i \pi \dsl q^f \theta(q_z^f)\delta(q_0^2-(q_z^f)^2).\label{ie}
\ee

The two conventions differ by the last term on this equation. However, this
term, as all matching terms of this type, do cancel among the different factors
in any of the eqs. (\ref{prol}), (\ref{pror}) and (\ref{propsym}).  This
cancellation is in fact the same phenomenon as the cancellation of the
residues corresponding to ``ambiguous $i\epsilon$'s'' that we discussed after
eqs. (\ref{eleft}), (\ref{eright}) and (\ref{ecross}). We are thus lead to the
striking conclusion that {\it the ``time'' and the ``space'' $i \epsilon$
conventions are strictly equivalent in eqs. (\ref{prol}), (\ref{pror}) and
(\ref{propsym})}.

This equivalence is physically understandable. It means that all the poles for
 which the two conventions give conflicting ``$i\epsilon$'' signs have a
 vanishing residue: they correspond to physically non acceptable asymptotic
 states. For instance, consider a massless pole with a positive $q_z^f$ in
 eq. (\ref{ie}). Massless asymptotic states can only exist in the unbroken
 phase, i.e. on the left side of the wall. A final state in the unbroken phase
 must necessarily move away from the wall, i.e. with a negative $q_z^f$. It is
 then physically  welcome that the residue vanishes when $q_z^f$ is
 positive. All other apparent conflicts are resolved in a similar way.

We shall denote the first line of expressions (\ref{prol}), (\ref{pror}) and
(\ref{propsym}) as the ``non-homogeneous" part. The rest of the propagator in
any of its versions will be denoted ``homogeneous part", as the latter is a
solution of the Dirac equation without a source, contrary to the former. By
the same token, it is possible to verify that the advanced and retarded Green
functions in the presence of a wall differ from those for the usual Dirac
propagator by terms which vanish upon the action of the Dirac operator.

The presence of both massless and massive poles reflects the fermion sailing
between the unbroken and broken phases.  It is trivial to verify in
eqs. (\ref{prol}), (\ref{pror}) and (\ref{propsym}) that the usual
$i/\slash{p}$ Feynman propagator for a massless fermion is recovered. Notice
as well that, besides the usual $i\epsilon$ dependence which corresponds to
absorptive contributions for on-shell fermions, new imaginary phases appear
which are generated by the reflection coefficients of the fermion when $E^2 <
m^2$. This CP-even phases, present for on-shell and off-shell fermions, are
induced by the reflection matrix, eq.(\ref{R}).

For completeness we give as well the fermion propagator in the presence of the
wall for the 4-dimensional case, i.e., when $q_x$ and $q_y$ have not been
boosted to zero values:
$$
S(q^f,q^i)=-1/2\bigg\{
\frac{1}{q^f_z-q^i_z+i\epsilon}
 \left(\frac{1}{\dsl q^f}+\frac{1}{\dsl q^i}\right) -
 \frac{1}{q^f_z-q^i_z-i\epsilon}
 \left(\frac{1}{\dsl q^f-m}+\frac{1}{\dsl q^i-m}\right) +
\nn
$$\nobreak
$$
\frac{1}{\dsl q^f-m}\gamma_z \frac{1}{\dsl q^i}-
 \frac{1}{\dsl q^f}\gamma_z \frac{1}{\dsl q^i-m}-
\nn
$$\nobreak
\be
\frac{m}{\dsl q^f(\dsl q^f-m)}
 s\left[\frac{E\gamma_0-q_x\gamma_x-q_y\gamma_y+\frac{m s \gamma_z}{p_z+p_z'}
 (E\gamma_0-q_x\gamma_x-q_y\gamma_y+s\gamma_z)}{p_z}\right]
 \frac{m}{\dsl q^i(\dsl q^i-m)}\bigg\}
\label{prop4}
\ee
where $p_z=\sqrt{E^2-q_x^2-q_y^2}+i\epsilon$ and
$p'_z=\sqrt{E^2-q_x^2-q_y^2-m^2} +i\epsilon$.

During the (long) preparation of this manuscript, we became aware of
ref.\cite{funakubo} which derives the Dirac propagator in presence of a thick
wall and uses it to compute the CP-asymmetry in the reflection process. The
source of CP-violation they use is not a SM loop but rather the tree-level
axial quark mass in a 2 scalar model. Although this difference makes the
comparison between both works delicate, their results seem to support our
statements that increasing the wall thickness would further reduce the
asymmetry.

\seCtion{One loop computation.}
\label{secloop}

\begin{figure}[hbt]
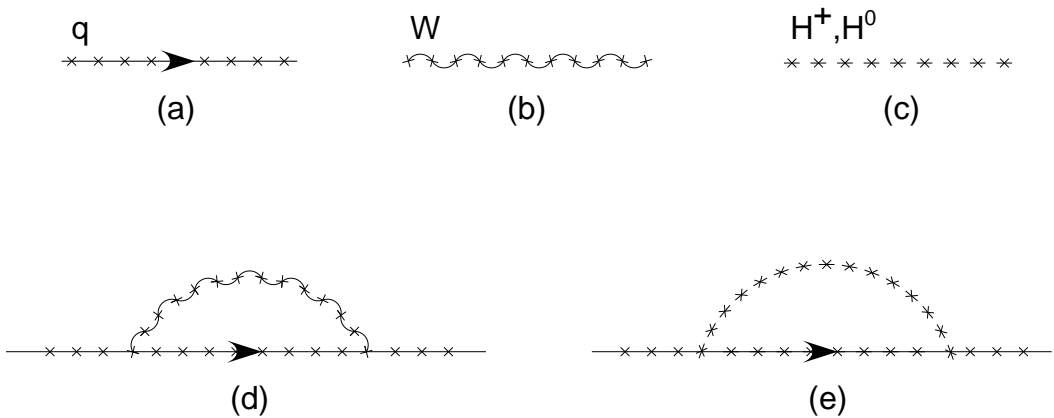
   
    \begin{center}
      \mbox{\psboxto(0.85\hsize;0pt){fullloop.ai}}
    \end{center}
    \caption{\it The kinky propagators for quarks, $W$'s and scalars can be
      assembled into loop diagrams. Vertices, being local, are unchanged in
      this formalism.}
    \protect\label{cloop}
\end{figure}

The propagators in the presence of the wall will be depicted by a line
screened by crosses (see Fig. \ref{cloop}), while the usual ones are drawn as
solid lines. Consider a one-loop transition between two external flavours, $i$
and and $f$, such as the ones in Fig. \ref{cloop}. The analogous diagrams with
full lines, i.e., with the usual Feynman propagators, are completely
renormalized away for on-shell external states. This will be changed by the
presence
of the wall, which acts on the diagram as a external source of momentum.

Let us review the properties expected in general, independently of the details
of a given calculation. The process under consideration describes an
asymptotically massless quark $i$, approaching the wall from the unbroken
phase, which is reflected into the same phase as an asymptotic massless $f$
quark. A flip of chirality has thus necessarily occurred. The flavour change
happens through one electroweak loop, and we will denote by $M$ the internal
fermion mass.

First of all, the amplitude should not vanish at
order $\alpha_W$. There is no symmetry reason known to us implying that the
coefficient of the CP-violating operator should be zero at this order.

Second, the amplitude should vanish at least as $m_im_f$ when the internal
loop is computed in just one of the two phases, and Lorentz covariance is
preserved. A priori this behaviour is not mandatory, as just one chirality
flip is needed, and an amplitude vanishing with masses as either $m_i$ or
would seem adequate. But there are more constraints: in the unbroken phase,
the computation of the internal loop should behave as
\be
   I_{\mu}(q_\mu)\propto \dsl q\,,
\label{gen}
\ee
due to Lorentz covariance. This loop is to
be inserted on the propagator in the presence of the wall. Imagine that
reflection has happened before the loop insertion (providing an $mi$ factor),
but not afterwards. The
outgoing state is then a plane wave verifying the Dirac equation for a
massless particle. The action of eq.(\ref{gen}) on it gives zero. In the
opposite case, when reflection happens only after the loop insertion, the
analogous argument for the incoming massless state holds as
well. Consequently, the action of the wall must be
present before and after the internal loop insertion, and a non trivial
dependence of the amplitude on both $m_i$ and $m_f$ results. The total effect
should go to zero as some positive power of both of them, with an odd overall
dependence on external masses, since chirality flips upon reflection. The
argument can be generalized to the situation where the broken phase is
considered inside the loop, the difference with eq. (\ref{gen}) being terms
independent of $\dsl q$, but proportional to the
external masses, as it will be exemplified below.

Third, it is possible to argue the type of GIM cancellations for the quark
masses inside the loop.  The relevant terms in the reflection amplitude
$A(i\rightarrow f)$ correspond to the interference of diagrams with two
different internal quark masses, $M$ and $M'$, to be summed upon.  Each
individual diagram can be written as\footnote{This dependence is in general
a complicate function. For transparency, only the behaviour for $M<M_W$ is
described here. As the internal loop is
IR convergent when $M\to0$, no pure $logM$ terms can appear.}
\be
A(i\rightarrow f) \propto F(M)=
a + \frac{M^2}{M_W^2}(b + c log\frac{M}{M_W})
 + \frac{M^4}{M_W^4}(d + e log\frac{M}{M_W})+.....\,,
\label{g1}
\ee
with $a,b,c,d,e$... depending only on external masses and
momenta.  Due to CKM
unitarity (see eq. (\ref{er1})), the CP observable is proportional to
$\sum_{M,M'} Im[F(M)F(M')^*]$. As each term in this sum is an antisymmetric
function of $M$ and $M'$, the squared terms in the development (\ref{g1}) of
the product cancel, while the cross-terms involving the constant piece $a$
fall out once the sum is performed. The lowest order contribution is thus
$\sim\sum\frac{M^2M'^2}{M_W^4}\log(M^2)$. In the practical example below, we
find $\frac{M^2M'^4}{M_W^6}\log M'^2$, i.e. a further internal mass
suppression.  It is important to notice that, whenever only the unbroken phase
is considered inside the electroweak loop, $c=d=e=0$, because the fermionic
mass dependence stems from pure Yukawa couplings. No antisymmetric function in
$M,M'$ is viable, and the effect should vanish at order $\alpha_W^2$ in total
rate. There is no reason, though, to neglect the broken phase, and we expect
an $O(\alpha_W^2)$ contribution.

Finally, the observable should be proportional to the interference of the
imaginary parts of the CKM couplings with the CP-even phases produced in the
reflection on the wall.

The above considerations apply as well for the thick wall scenario. Some
remarks specifically related to the wall thickness are appropriate, before
entering into the details of our computation. The internal electroweak loop is
in general a complicate function of $q^2$, and thus non-local. Its typical
``inverse-size" is of the order of the dimensional parameters involved which
means $\ge M_W$, as we will see. To neglect its non-local character is only
adequate for wall thicknesses $l$ larger than the loop ``size",
$l>>M_W^{-1}$. For a thin wall, $l\rightarrow 0$, a local approximation for the
internal loop is incorrect.  Arbitrary large particle momenta are relevant, and
it will be shown that indeed the non-local character of the internal loop is
important. The role played by the latter effects in our calculation suggests
that an even smaller result would follow in a more realistic thick wall
computation, $l>>M_W^{-1}$, where a local internal loop could be a good
approximation.

We will see that the above four general characteristics do appear in the toy
computation given below.

\subsection{One particle irreducible loop}
\label{secpi}

\begin{figure}[hbt]
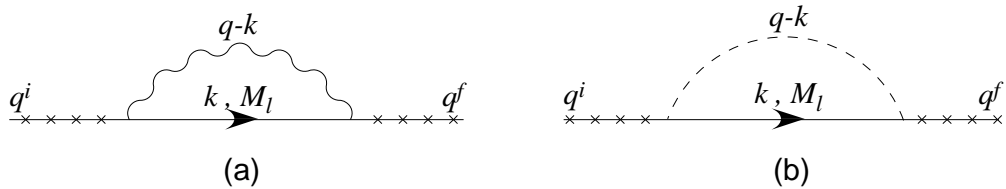
   
    \begin{center}
      \mbox{\psboxto(0.85\hsize;0pt){simploopt.ai}}
    \end{center}
    \caption{\it The subset of Feynmann diagrams used in this paper.}
    \protect\label{simploop}
\end{figure}

The complete computation of the diagrams in Fig. \ref{cloop} requires not only
the knowledge of the fermion propagator in the presence of the wall, but the
corresponding one for gauge and Higgs bosons as well. They can be
obtained in complete analogy with the computations of the
previous section.  However, they need rather lengthy calculations. Our goal
here is rather to get a feeling of how the different building blocks work
together than to solve exactly a problem which anyhow is academic to a large
extent.  We have chosen to simplify our task, leaving the full calculation for
a forthcoming publication. What we compute in fact are the
diagrams in Fig. \ref{simploop}, with the internal loop computed in the broken
phase.  Notice that a similar choice has been made in recent publications which
try to solve the problem at finite temperature\cite{shapo}, where only the
unbroken phase was considered inside the electroweak loops.

Let us then consider the broken phase inside the loop. In the 't Hooft Feynman
gauge the integral to perform can be written in general as
\be
I(q_{\mu})= c \int^{\Lambda} d^4 k \frac {\dsl k a +b} {\left ( (q-k)^2-M_W^2
\right)(k^2-M_l^2)},
\label{iq}
\ee
where $q_\mu$ is the external 4-momentum and $a,b$ and $c$ are
$k$-independent. $\Lambda$ indicates that the
integral has been regularized someway, for instance by dimensional
regularization.

 We choose to substract at $q^2=0$. Because Lorentz
covariance is preserved at $T=0$, the result of the substraction will have the
general form
\be
I^{subst.}(q_{\mu}) = \slash{q} A^{subst.}(q^2) + B^{subst.}(q^2)\,,
\label{iqs}
\ee
where A and B are matrices with $1$ and/or $\gamma_5$ components,
$A^{subst.}(q^2)= A(q^2)-A(0)$ and $B^{subst.}(q^2)= B(q^2)-B(0)$. Finally,
\bea
I^{subst.}(q_{\mu}) =
\pi^2 c \int_0^1 dx \int_0^1 dy (\slash{q} a x + b) \frac {q^2 x(1-x)}
{q^2xy(1-x)-M_W^2 x - M_l^2 (1-x)}.
\label{iqs2}
\eea

Notice that with, the above choice of substraction, the internal loop is
completely renormalized away for an on-shell massless fermion which suffers no
interactions with the wall.  Concentrate for the moment in the case $i,f= d,s$
or $b$. The scalar-exchange contribution is then
\bea
I^{subst.}_{scalar}(q_{\mu}) =
\frac {g^2} {2M_W^2}
\frac {\pi^2} {(2\pi)^4}
K^*_{lf}K_{li}
 \int_0^1 dx \int_0^1 dy  \frac {q^2 x(1-x)}{q^2xy(1-x)-M_W^2 x - M_l^2 (1-x)}
\nn \\
\{ x \dsl q (m_i m_f R + M_l^2 L) - M_l^2(m_fL+m_iR) \},\nn\\
\label{iqscal}
\eea
where $L$ and $R$ denote the chiral projectors, $L=(1-\gamma_5)/2$,
$R=(1+\gamma_5)/2$. $K$ describes the CKM couplings of the quarks and $M_l$ is
the mass of the internal quark of $l$-th generation.

The  W-exchange contribution to the substracted internal loop is
\bea
I^{subst.}_{W}(q_{\mu}) =
g^2
\frac {\pi^2} {(2\pi)^4}
K^*_{lf}K_{li}
 \int_0^1 dx \int_0^1 dy  \frac {q^2 x(1-x)}{q^2xy(1-x)-M_W^2 x - M_l^2 (1-x)}
{x\slash{q} L}.\nn\\
\label{iqscal2}
\eea

The above represent only one contribution among many that should be considered
in the full calculation.  When this internal loop is inserted below on our
propagator, gauge invariance is preserved, as the fermion interaction with the
wall is of Yukawa type and thus $SU(3)\times SU(2)\times U(1)$ gauge
invariant. In the full theory, the substraction procedure will be more
complicated, though\footnote{What has been done above is close to the complete
result in a limited extension of the standard model, with two Higgs's, one
($\Phi_d$) coupled to down quarks and the other ($\Phi_u$) to up quarks. We
assume the following expectation values: $<\Phi_d(z)>=v_d \theta(z)$ and
$<\Phi_u(z)>=v_u$, with $v_d\ll v_u$. In this model the equation
(\ref{iqscal2}) turns out to become the exact contribution from W exchange,
while eq. (\ref{iqscal}), or rather its Fourier transform, is modified by the
following substitution: $m_i\ra m_i\theta(z_i),\, m_f\ra m_f \theta(z_f)$,
where $z_i$ ($z_f$) is the coordinate of the incoming (outgoing) vertex of the
loop.  An analysis of the changes in the computations runs parallel to the one
performed above, but it does not seem pertinent to pursue this model
here.}. Our aim in a toy computation is to get an idea of how the ingredients
of CP vilolation, GIM mechanism, and wall reflection combine together, and to
settle useful calculational tools for this type of problem.

\subsection{Loop insertion in the propagator}
\label{seclp}
The diagrams in Fig. \ref{simploop} are computed inserting the internal loops
given above on the fermion propagator in the presence of the wall, derived in
the preceding section. The desired amplitude should describe the transition of
an incoming massless flavour $i$ into an outgoing massless flavour $f$. The
selection of these asymptotic states is obtained projecting onto the massless
poles of the propagator, for the 4-momenta appearing at both extremes of
Fig. \ref{simploop}.

The relevant contributions of Fig.(\ref{simploop}) are then those terms which
contain both poles in $1/\dsl q^i$ and $1/\dsl q^f$. It can be shown that the
non-homogeneous part of the fermion propagator in the presence of the wall,
gives no contribution of this type. That is, among the four possible
combinations of homogeneous and non homogeneous on both sides of the internal
loop, only the homogeneous-internal loop-homogeneous contributes, yielding for
a positive energy particle:
\bea
A(i\rightarrow f)= \int \frac{dq_z}{2\pi}
\frac{1}{\dsl q^f}
 \left[\frac{1+ \alpha_z}{2}\right]
 \left[1-\frac{m_f  \gamma_0}{E+{p'_z}^f}\right]  \gamma_0
 \frac{m_f}{\dsl q(\dsl q-m_f)}\nn \\
\left[I^{subst.}_{W}(q_{\mu})+I^{subst.}_{scalar}(q_{\mu})\right]
\frac{m_i}{\dsl q(\dsl q-m_i)}
 \left[1-\frac{m_i  \gamma_0}{E+{p'_z}^i}\right]
 \left[\frac{1-  \alpha_z}{2}\right]  \gamma_0
 \frac{1}{\dsl q^i},
\label{lhom}
\eea
where
\be
{p'_z}^i= +\sqrt{E^2-m_i^2}+i\epsilon
\label{p0i}
\ee
and
\be
{p'_z}^f= +\sqrt{E^2-m_f^2}+i\epsilon .
\label{p0f}
\ee
The $q_z$ integration reflects  the arbitrary off-shellness of the quarks
upon reflection on the wall. The relevant integrals are the following ones:
\bea
\lefteqn{\int \frac{dq_z}{2\pi}
\frac{1}{q^2-m_i^2}
\frac{1}{q^2-m_f^2}
\frac{1}{q^2-{\widetilde{M}}_l^2}=}\nn \\
&&\frac{i}{2(m_i^2-{\widetilde M}_l^2)(m_f^2-{\widetilde M}_l^2)}
\bigg [\frac{1}{{p'_z}^i{p'_z}^f}
({p'_z}^f + \frac{{\widetilde M}_l^2-m_f^2}{{p'_z}^i+{p'_z}^f})-
\frac{1}{\widetilde Q_0 }\bigg ]
\label{dqz1}
\eea
and

\bea \lefteqn{\int \frac{dq_z}{2\pi} q^2
\frac{1}{q^2-m_i^2}
\frac{1}{q^2-m_f^2}
\frac{1}{q^2-{\widetilde{M}}_l^2}=} \nn \\
& &\frac{i}{2(m_i^2-{\widetilde M}_l^2)(m_f^2-{\widetilde M}_l^2)}
\bigg [\frac{1}{{p'_z}^i{p'_z}^f}
({p'_z}^f \widetilde M_l^2+ m_i^2\frac{{\widetilde
M}_l^2-m_f^2}{{p'_z}^i+{p'_z}^f})-
\frac{\widetilde M_l^2}{\widetilde Q_0 }\bigg ],
\label{dqz2}
\eea
where
\be
\widetilde M_l^2=\frac{xM_W+(1-x)M_l^2}{xy(1-x)},
\label{mtilde}
\ee
\be
\widetilde Q_0=\sqrt{E^2-\widetilde M_l^2}+i\epsilon\,.
\label{qtilde}
\ee
After some lenghty algebraic manipulations, the resulting amplitude for an
incoming left asymptotic quark can be written as
\bea
A(i_R\rightarrow f_L)&=&
\frac{g^2}{2} \frac{\pi^2}{(2\pi)^4}
K_{li}K^*_{lf} \int_0^1 dx \int_0^1 \frac{dy}{y}
\frac{i}{(m_i^2-{\widetilde M}_l^2)(m_f^2-{\widetilde M}_l^2)}\nn\\
&&\frac{1}{\dsl q^f} m_i m_f^2
\left [
\bigg (\frac{1}{{p'_z}^i{p'_z}^f}
({p'_z}^i \widetilde M_l^2+ m_f^2\frac{{\widetilde
M}_l^2-m_i^2}{{p'_z}^i+{p'_z}^f})-
\frac{\widetilde M_l^2}{\widetilde Q_0}\bigg )
  [x\tilde{a}+\tilde{b}]+\right.\nn\\
&&\left.\bigg(\frac{1}{{p'_z}^i{p'_z}^f} ({p'_z}^i + \frac{{\widetilde
M}_l^2-m_i^2}{{p'_z}^i+{p'_z}^f})-
\frac{1}{\widetilde Q_0 }\bigg )[x\tilde{c}+\tilde{d}]
 \right ]
\frac{1+\alpha_z}{4} R \frac{1}{\dsl q^i},
\label{lws}
\eea
where
\bea
\tilde{a}&=& (2+\frac{M_l^2}{M_W^2})(1-\frac{{p'_z}^i}{E+{p'_z}^f})+
\frac{m_i^2}{M_W^2}(1-\frac{{p'_z}^f}{E+{p'_z}^i}),\\
\tilde{b}&=& -\frac{M_l^2}{M_W^2}(1+\frac{E-{p'_z}^i}{E+{p'_z}^f}),\\
 \tilde{c}&=& -m_i^2\left[\frac{E}{E+{p'_z}^i}(2+\frac{M_l^2}{M_W^2})+
\frac{m_f^2}{M_W^2}\frac{E}{E+{p'_z}^f}\right],\\
 \tilde{d}&=& \frac{M_l^2}{M_W^2}(E^2-{p'_z}^i{p'_z}^f)
(1+\frac{E-{p'_z}^i}{E+{p'_z}^f}).
\label{atod}
\eea

Notice that $\widetilde Q_0$, obtained from the convolution of the electroweak
loop with the free fermion propagator in the presence of the wall, is in
general complex, and will be one of the relevant CP-even phases which will
interfere with the CKM complex couplings. This new phase stems from the
non-local character of the internal electroweak loop, and would have
disappeared in a linear approximation. Other contributions of the same type
should be present in a full calculation with the wall accounted for inside the
loop, and although cancellations are possible, they are not mandatory by any
symmetry argument.

For an incident left-handed particle, the result is obtained from
eqs. (\ref{lws}) to (\ref{atod}) replacing everywhere $m_i$ by $m_f$ and
viceversa (and consequently ${p'_z}^i$ by ${p'_z}^f$). The CKM couplings
remain identical.

Consider transitions between charge $-1/3$ quarks. Concentrate as well
on quark energies below or equal to $M_W$. The approximation
$m_i^2,m_f^2,E^2 < \widetilde M_l^2$ is then justified, and eq. (\ref{lws})
simplifies to
\bea
A(i_R\rightarrow f_L)&=&
\frac{g^2}{2} \frac{\pi^2}{(2\pi)^4}
K_{li}K^*_{lf} \int_0^1 dx \int_0^1 \frac{dy}{y}
\frac{i}{\widetilde M_l^4}\nn\\
&&\frac{1}{\dsl q^f} m_i m_f^2
\left [
\bigg (\frac{1}{{p'_z}^i{p'_z}^f}
({p'_z}^i + \frac{m_f^2}{{p'_z}^i+{p'_z}^f})-
\frac{1}{i\widetilde M_l }\bigg )\widetilde M_l^2[xa+b]+\right.\nn\\
&&\left.\bigg (\frac{1}{{p'_z}^i{p'_z}^f}
({p'_z}^i + \frac{{\widetilde M}_l^2}{{p'_z}^i+{p'_z}^f})-
\frac{1}{i\widetilde M_l }\bigg )[xc+d]
 \right ]
\frac{1+\alpha_z}{4} R \frac{1}{\dsl q^i},
\label{lws2}
\eea
where $a,b,c,d,$ are given by eq. (\ref{atod}), neglecting  terms in
$m_i^2/M_W^2$,$m_f^2/M_W^2$.

The amplitude can be rewritten as
\bea
A(i_R\rightarrow f_L)&=&
-i\frac{g^2}{2} \frac{\pi^2}{(2\pi)^4}
K_{li}K^*_{lf}
\frac{1}{\dsl q^f} \frac{m_i m_f}{M_W^2}
\,F(M)\,
\frac{1+\alpha_z}{4} L \frac{1}{\dsl q^i}.
\label{lws3}
\eea
The function $F(M)$ is defined as follows:
\bea
F(M)&=& m_f\frac{1}{{p'_z}^i+{p'_z}^f}(1+\frac{E-{p'_z}^i}{E+{p'_z}^f})
  [I_1- I_2]
\\
&&+i
\frac{m_f}{M_W}\left[(1+\frac{E-{p'_z}^i}{E+{p'_z}^f}) I_3-
  (1-\frac{{p'_z}^i}{E+{p'_z}^f}) I_4\right],
\label{ef}
\eea
and the integrals $I_{1,2,3,4}$ are
\be
I_n(M)=\delta_n\int_0^1dx\left[\frac{x(1-x)}{x+
\frac{M^2}{M_W^2}(1-x)}\right]^{\gamma_n} x^{\beta_n}\,,\nn \\
\label{int}
\ee
with $\delta_1=3$, $\delta_3=M^2/M_W^2$,
$\delta_2=3\delta_4/2=2+M^2/M_W^2$, $\beta_n=(0,1,0,1)$ and
$\gamma_n=(1,1,3/2,3/2)$.

Consider $i\ne f$, i.e., a flavour changing transition. The CP-violating
observable is proportional to

\bea
\lefteqn{\sum_{l,l'}[|\tilde{A}(i_R\rightarrow f_L)|^2
 -|\tilde{A}(\bar i_L\rightarrow \bar f_R)|^2]=
[\frac{g^2}{4} \frac{\pi^2}{(2\pi)^4}]^2(\frac{m_im_f}{M_W^2})^2}\nn\\
&&(-2)\sum_{l,l'}Im\bigg[K_{li}K^*_{lf}(K_{l'i}K^*_{l'f})^*\bigg]
 Im\bigg[ F(M_l)F^*(M_{l'})\bigg],
\label{er1}
\eea
where $\tilde{A}(i\ra f)$ is obtained from $A(i\ra f)$ taking the residues of
$1/{\dsl q^i}$ and $1/{\dsl q^f}$ at $q_0=q^i_z=-q^f_z$. $\tilde{A}(i\ra f)$
corresponds to the matrix
element of the effective Hamiltonian between states $|i>$ and $<f|$, in a
normalization that will be specified in the next subsection.

Eq. (\ref{er1}) shows explicitly the interference between the CP-odd phases of
the CKM couplings $K$ and the CP-even phases of $F$. The latter can have
two different origins: either $p_z'$ when $i$ or $f$ are totally reflected, or
$\widetilde Q_0$, which leaves a trace of the non-locality of the internal loop
in
the $i$ factor before $I_3$ and $I_4$.

It is well known that
\be
Im\bigg[K_{li}K^*_{lf}(K_{l'i}K^*_{l'f})^*\bigg]=
\eta J\,,\label{imk}
\ee
where $\eta=1$ if $f-i=l'-l \pmod 3$, $\eta=-1$ when $f-i\ne l'-l \pmod 3$ and
where J is twice the area of the CKM unitarity triangle,
$J=c_1c_2c_3s_1^2s_2s_3s_\delta$. Let us now consider in the complex plane the
triangle spanned by the three numbers
$F(M_l), l=u,c,t$. Let $B$ be the area of this triangle
with a sign +1 (-1) for a cyclic clockwise (anti-clockwise) ordering of
$u,c,t$ around the triangle. Then
\be
\sum_{l,l'}
Im\bigg[K_{li}K^*_{lf}(K_{l'i}K^*_{l'f})^*\bigg]
Im\bigg[ F(M_l) F^*(M_{l'})\bigg]
= 8JB.\label{tri}\ee

This expression exhibits clearly the GIM mechanism for the internal flavours,
since the area vanishes obviously whenever two points coincide. Additional
vanishing happens when three points are aligned.  The function $F$
depends both on the internal and external
masses as can be seen from eq. (\ref{ef}). In
order to exhibit the GIM mechanism for the external masses as well,
$Im\bigg[F(M_l)F^*(M_{l'})\bigg]$ can be expanded,
leading to:
\bea
\sum_{l,l'}[|\tilde{A}(i_R\rightarrow f_L)|^2
 -|\tilde{A}(\bar i_L\rightarrow \bar f_R)|^2]=
[\frac{g^2}{4} \frac{\pi^2}{(2\pi)^4}]^2(-c_1c_2c_3s_1^2s_2s_3s_\delta)\nn\\
\sum_{l,l'=l+1 \pmod 3}(-2)\sum_{jk} S_{jk}(M_l,M_{l'})b_{jk}(E,m_i,m_f).
\label{er2}
\eea
In this expression, the function $S$ derives from the integrals $I_{1,2,3,4}$
in
eq. (\ref{int}), and contains the dependence on the masses for the quarks
inside the loop,
\be
S_{jk}(M,M')=I_j(M)I_k(M')-I_j(M')I_k(M),
\label{es}
\ee
and the $b_{jk}$ are antisymmetric functions in $j,k$. They contain the
dependence on the masses of the incoming and outgoing quarks. In the
approximation used from eq. (\ref{lws2}) on, they are equal to
\bea
b_{12}(E,m_i,m_f)&=&0,\\
b_{13}(E,m_i,m_f)&=&-(\frac{m_im_f}{M_W^2})^2
\frac{m_f^2}{M_W}
\frac{|2E+{p'_z}^f-{p'_z}^i|^2}{|E+{p'_z}^f|^2}
Re\bigg(\frac{1}{{p'_z}^i+{p'_z}^f}\bigg),\\
b_{23}(E,m_i,m_f)&=&-b_{13}(E,m_i,m_f),\\
b_{14}(E,m_i,m_f)&=&+(\frac{m_im_f}{M_W^2})^2
\frac{m_f^2}{M_W}\frac{1}{|E+{p'_z}^f|^2}\times\nn\\
&&Re\bigg(\frac{(2E+{p'_z}^f-{p'_z}^i)(E+{{p'_z}^f}^*-{{p'_z}^i}^*}
  {{p'_z}^i+{p'_z}^f}\bigg),\\
b_{24}(E,m_i,m_f)&=&-b_{14}(E,m_i,m_f),\\
b_{34}(E,m_i,m_f)&=&(\frac{m_im_f}{M_W^2})^2
\frac{m_f^2}{M_W^2}\frac{E}{|E+{p'_z}^f|^2}Im\bigg ({p'_z}^f-{p'_z}^i\bigg ).
\label{bjk}
\eea

For each $j,k$ pair, $\{I_j(M),I_k(M)\}$ can be seen as the coordinates of
the point $M$ in the $(j,k)$-plane. Then the points $(m_u,m_c,m_t)$ span a
triangle in this plane whose area $1/2\sum_{l,l'=l+1\pmod 3}
S_{jk}(M_l,M_{l'})$ exhibits
again the GIM mechanism for internal
masses. Furthermore, because
$Im\bigg[(K_{i,l}K_{l,f})(K_{i,l'}K_{l',f})^*\bigg]$ is antisymmetric in the
exchange of $i$ and $f$, the sum over external flavours is proportional to
\be
\propto \sum_{i,f=i+1 \pmod 3} (b_{jk}(E,m_i,m_f)-b_{jk}(E,m_f,m_i)),
\ee
which shows
explicitly the GIM mechanism operating on external quarks. This pattern should
be present in a complete computation, with pertinent modifications in the
functions $S_{jk}, b_{jk}$.

All $b_{jk}$ but $b_{12}$ require the existence of the CP-even phase in
$\widetilde Q_0$ found above, and depend as well on the complex reflection
behaviour of the initial and final quarks, contained in ${p'_z}^i$ and
${p'_z}^f$. $b_{12}$ depends only on the latter, and could be $\ne 0$ only
when $m_i$,$m_f$ and $E$ are not neglected in front of $M_W$.

Note that all $b_{jk}$ vanish when either $m_i$ or $m_f$ go to zero,
respecting the second point of the general behaviour announced at the
beginning of this section. In each sector of quark charges ($-1/3$ or +$2/3$)
the heavier masses will dominate the effect, as expected from intuition. Note
that a two-threshold structure is present, corresponding to $m_i$, $m_f$.

 In our numerical results we use the exact values of the fonctions
$S_{jk}(M,M')$. It is instructive, though, to show a fit for the particular
case $M,M'<<M_W$, an appropriate expansion for all quarks but the top,

\be
S_{1,2}(M,M')\rightarrow\frac{M^2M'^2}{M_W
^4}\left[\frac{M^2}{M_W^2}log\frac{M^2}{M_W^2}-\frac{M'^2}{M_W^2}
log\frac{M'^2}{M_W^2}\right],
\label{sap}
\ee
and the ratio of the remaining $S_{j,k}(M,M')$ to the result in eq.
(\ref{sap})  is given by $S_{1,3}/S_{1,2}\rightarrow -\frac{7}{15}$,
$S_{1,4}/S_{1,2}\rightarrow +1.5$,
$S_{2,3}/S_{1,2}\rightarrow -\frac{1}{30}$,
$S_{2,4}/S_{1,2}\rightarrow -\frac{1}{4}$,
and $S_{3,4}/S_{1,2}\rightarrow +\frac{1}{6}$, in the same limit. This
behaviour is consistent with the third point of the general expectations
developed at the beginning of this section.

\subsection{Transition amplitude in terms of wave functions}
\label{seclwf}

The computation of the amplitude described in the previous subsection was
obtained by projecting over the massless poles of the propagator in the
presence of the wall, once a flavour-changing electroweak loop was inserted.

The result should be equivalent to the more direct one in terms of incoming
and outgoing wave functions (wave packets), as given by eq. (\ref{wp3}), with
no need to consider the full propagator in the presence of the wall derived in
Section \ref{secprop}, as only the usual propagators were used inside
loops. To prove that such is the case, consider the incoming wave function
$\widetilde\psi^{inc}_{n^+}$ in eq. (\ref{fpsin}). The following
identities\footnote{recall that $p_z=\sqrt{E^2}=q_0$ in this
case.}
\bea
\frac{1}{\dsl q}\gamma_0&=&\frac{1}{p_z-q_z+i\epsilon}\frac{1+{\alpha}_z}{2}+
\frac{1}{p_z+q_z+i\epsilon}\frac{1-{\alpha}_z}{2},\label{identunbr}\\
\frac{1}{\dsl q-m}\gamma_0&=&\frac{1}{p_z'}
\bigg[\frac{1}{2}\frac{p_z+p_z'{\alpha}_z+m\gamma_0}{p'_z-q_z+i\epsilon}+
\frac{1}{2}\frac{p_z-p_z'{\alpha}_z+m\gamma_0}{p'_z+q_z+i\epsilon}\bigg],
\label{identbr}
\eea
lead to the equivalent expression
\bea
\widetilde\psi^{inc}_{n^+}(q_z)=\bigg[-\frac{1}{2\pi i}
\frac{m}{\dsl q(\dsl q-m)}
(1-\frac{m\gamma_0}{p_z+p_z'})
\gamma_0+\delta (p_z-q_z)\bigg]
\frac{1+{\alpha}_z}{2}u_h(\pinc).
\label{fpsin2}
\eea
In the neighbourhood of the massless poles of the propagator of a given quark,
\def\arti{ {\,\,\lower .8ex\hbox {$\longrightarrow
 \atop {q^i_z\rightarrow E}$}\,\,}}
\def\artf{ {\,\,\lower .8ex\hbox {$\longrightarrow
 \atop {q^f_z\rightarrow -E}$}\,\,}}
\bea S(q^f,q^i)&\arti&
    2\pi i \sum_{h}\frac{\widetilde\psi^{inc}_{n+}(q^f_z)\,u_h(\pinc)\dagger}
   {E-q^i_z+i\epsilon},\\
S(q^f,q^i)&\artf&
2\pi i \sum_{h}\frac{u_h(\pout)\,\widetilde\psi^{out}_{n+}(q^i_z)\dagger}
{E+q^f_z+i\epsilon}.
\label{ay1}\eea

In eq. (\ref{fpsin2}) the first term corresponds to the residue at the
massless pole of the ``homogeneous" part of the propagator, while the term in
$\delta(p-q_z)$ corresponds to the residue at the same pole of the ``non
homogeneous" part of the propagator. For the particular choice of substraction
(at $q^2=0$) in subsection \ref{secpi}, the latter obviously gives no
contribution.  The conclusion found in the previous subsection, namely that
only the homogeneous parts of the propagator participated in the amplitude
under consideration, gets thus a simple interpretation in terms of the wave
function contributions to $<\psi^{out}_n| H_{int}| \psi^{inc}_n>$. And the
equivalence of both methods to compute the amplitude is proved.

\seCtion{CP Asymmetry}
\label{secCP}

\subsection{Reflection probability}
\label{secrefl}

The task is to estimate the probability of reflecting a quark of flavour $i$
into another flavour $f$, for a given flux of incoming particles
(antiparticles), i.e. the reflected flux per unit incoming flux. Let us
consider a finite box of length $L_z$ and unit
section in the plane $x-y$, which incorporates the wall. For simplicity, we
will
ignore in the following the $x$ and $y$ directions. We normalize the wave
functions $\psi^{inc}$ and $\psi^{out}$ in that box.  Denote the matrix
element of the interaction Hamiltonian, generated by
the loop between the latter states, by
\be
<\psi^{out}_{n_f}|H_{int}|\psi^{inc}_{n_i}> =
  \frac {\tilde{A}({i\ra f})}{L_z}\,,\label{hfi}
\ee
where $\tilde{A}(i\ra f)$ is dimensionless, see eqs.(\ref{er1}),
(\ref{er2}). For any time interval $\cal{T}$ such that $\tilde{A}(i\ra
f)\cal{T}/L_z<<1$,
\be
<\psi^{out}_{n_f}|e^{-iHT}|\psi^{inc}_{n_i}> \sim
e^{-iE\cal{T}}\frac{\tilde{A}(i\ra
  f)}{L_z}\, \int^{\cal{T}/2}_{-\cal{T} /2}dt e^{i\frac{1}{2}(E_f-E_i)t}
\label{hfi2}
\ee
whose modulus squared gives the reflection probability per particle
($E=(E_f+E_i)/2$ in this equation). Taking now the usual
limit:
\def\arTi{ {\,\,\lower .8ex\hbox {$\longrightarrow
 \atop \cal{T} \rightarrow \infty$}\,\,}}
\be
\left[\int^{\cal{T}/2}_{-\cal{T}/2}dt e^{i\frac{1}{2}(E_f-E_i)t} \right]^2
\arTi \cal{T} \,2\pi
\delta(E_f-E_i).\label{T}
\ee
For an incoming flux $\Phi$ per unit time of particles with
velocity $v$, the number of
particles in the volume will be $\Phi\,v\, L_z$.  The reflection probability
per unit time and unit flux is then:
\be
P(i\rightarrow f)\sim \sum_{p^f_z}\frac{\vert \tilde{A}(i\ra f)\vert^2}{L_z}
\,v\, 2\pi\delta(E_f-E_i). \label{prob}
\ee
In the infinite volume limit, $\sum_{p^f_z}/L_z \rightarrow \int dp^f_z
/(2\pi)$, and with $\int dp^f_z \delta(E_f-E_i)=1/v$, it follows that
\be
P(i\rightarrow f)=  \vert \tilde{A}(i\ra f)\vert^2.
\label{prob2}
\ee
The number of particles reflected per unit time (or reflected flux) is
$\Phi_r=|\tilde A (i\ra f)|^2 \Phi_i$, and thus related to the incoming flux
through the previously computed $|\tilde A|^2$.

Assume that the probability of the incoming state $i$ in the unbroken phase is
given by a distribution $\rho(i)$, which depends in general on the flavour,
chirality and energy of $i$. For instance, at finite temperature close to
equilibrium, this distribution will be close to the Fermi-Dirac one, with
proper modification of the dispersion relation between momentum and energy. We
define useful CP asymmetries by:
\be
\Delta_{CP}=\frac{\sum_{i,f}\rho(i)(P(i \rightarrow f)
- P(\bar i\rightarrow \bar f))}
{\sum_{i'}2\rho(i')},\label{CPas}
\ee
where the sum over $i,f$ in the numerator implies the sum
over the energies
and flavours for the isosinglet quarks (or the isodoublet
quarks)\footnote{The reason not to add in the asymmetry the isosinglet and
isodoublet quarks is the TCP' symmetry, eq. (\ref{asym1}), which implies that
their incoming probability $\rho(i)$ are equal. The eventual effect of
sphalerons to transform the CP asymmetry generated by the wall into a baryon
number asymmetry, is anyhow different for doublets and singlets. We thus need
to distinguish the asymmetry for doublets and singlets.}, while the sum $i'$
in the denominator acts over flavour and
helicities
and the factor 2 stems from
summing over quarks and antiquarks. In the case of zero temperature, if we
assume an equal incoming probability for all the flavours and helicities, and
considering the asymmetry for one given energy (for instance the most
favorable one) the formula (\ref{CPas}) simplifies to:
\be
\Delta_{CP}(\omega)^{T=0}=\sum_{i,f}\Delta_{CP}(\omega)^{T=0}_{i,f}
,\label{CPas2}\ee
 where
 \be \Delta_{CP}(\omega)^{T=0}_{i,f}=\frac{P(i\rightarrow f)
- P(\bar i\rightarrow \bar f)}{4N_{fl}}\label{CPas2if}
\ee
is the contribution to the asymmetry of a given pair of flavours. $N_{fl}$
denotes the numbers of flavours, 6. The colour factor has been obviated as it
will finally cancel between nominator and denominator.

\subsection{Numerical results}
\label{secns}

 Consider first the reflection asymmetries in the ``down" quark sector. The
numerical contribution of the functions $S_{jk}(M_l,M_{l'})$, which contain
the dependence on the internal quark masses ($l,l'=u,c$ and $t$), is the
following one
\be
\begin{array}{ll}
 \sum S_{1,2}(M_l,M_{l'})=3.82\,10^{-5}
&\sum S_{1,3}(M_l,M_{l'})=-1.84\,10^{-5}\\
 \sum S_{1,4}(M_l,M_{l'})=\,3.68\,10^{-5}
&\sum S_{2,3}(M_l,M_{l'})=-9.96\,10^{-7}\\
 \sum S_{2,4}(M_l,M_{l'})=\,5.74\,10^{-7}
&\sum S_{3,4}(M_l,M_{l'})=\,6.81\,10^{-7},
\end{array}
\label{sapn}
\ee
where the sums on $l,l'$ are constrained by the relation $l'=l+1 \pmod 3$.

\begin{figure}[hbt]
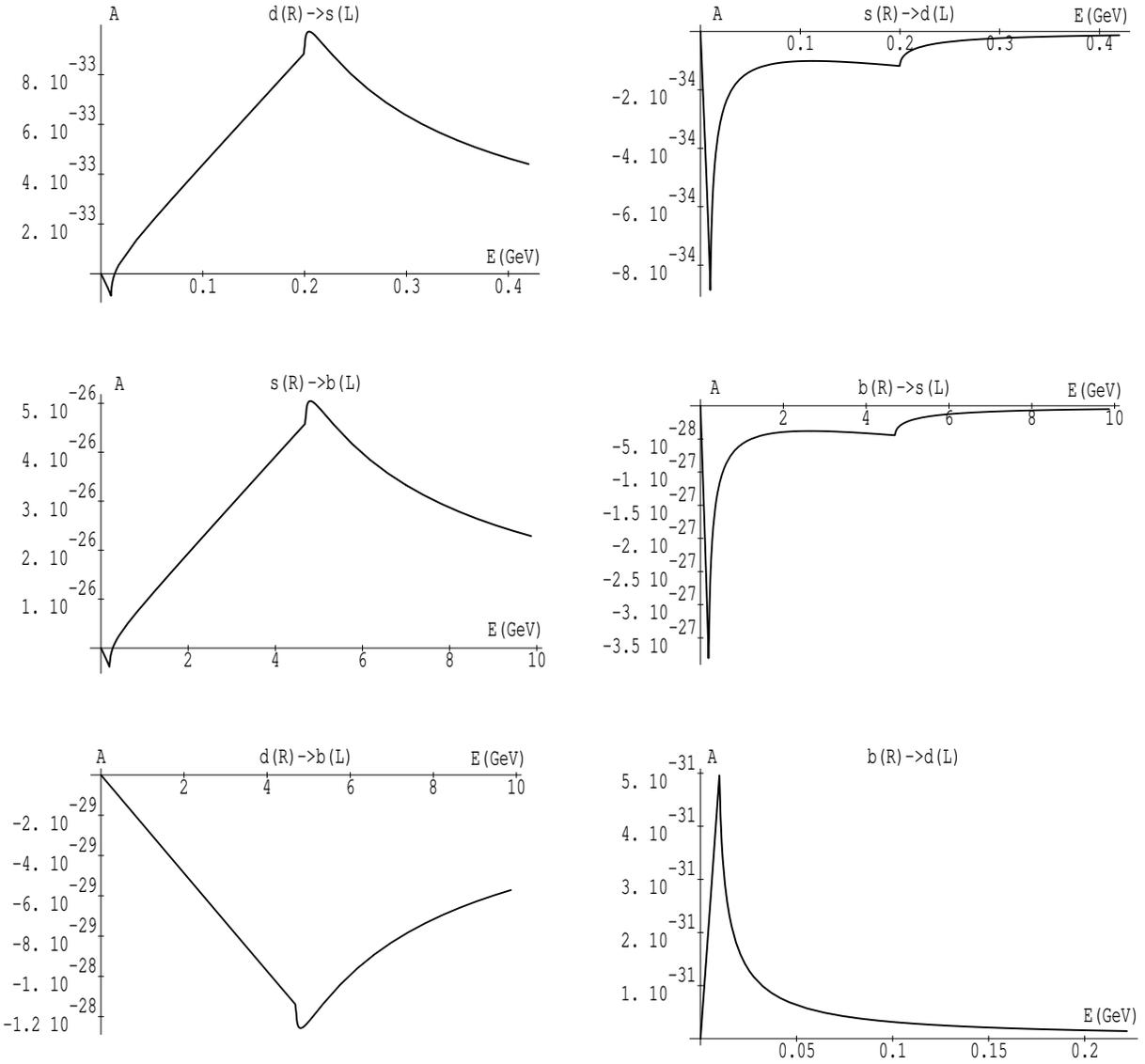
   
    \begin{center}
      \mbox{\psboxto(\hsize;0pt){asydown.ai}}
    \end{center}
    \caption{\it Asymmetries for incoming down quarks, as a function of
energy.}
    \protect\label{asydown}
\end{figure}

The following values have been used for the particle masses: $M_W=80.22$GeV,
$m_u=5$MeV, $m_d=10$MeV, $m_s=200$MeV, $m_c=1.3$GeV, $m_b=4.7$GeV,
$m_t=120$GeV. Our numerical results are presented in Fig. \ref{asydown}. As
expected the heavier quark masses dominate. This is well understood from the
analytical behaviour eq. (\ref{bjk}). The dominant contribution for charge
$-1/3$ quarks is thus given by the $s-b$ pair. The two spikes in the figures
correspond to the thresholds for the external quark masses involved. Indeed,
the analytical formulae in eq. (\ref{bjk}) display such a
structure, with for instance in the case $s_R\rightarrow b_L$:
\be
\begin{array}{lll}
  b_{12}=0\,,
& b_{14}=-(\frac{m_sm_b}{M_W^2})^2\frac{m_s}{M_W}\,,
 & \hbox{at } E=m_s
\\b_{13}\sim -\left(\frac{m_sm_b}{M_W^2}\right)^2\frac{m_b}{M_W}\,
& b_{14}\sim +\left(\frac{m_sm_b}{M_W^2}\right)^2\frac{m_b}{M_W}
   \frac{1}{2}\frac{m_s^2}{m_b^2}\,,
 & \hbox{at } E=m_b
\end{array}
\label{stob}
\ee
while for the complementary transition $b_R\rightarrow s_L$,
\be
\begin{array}{lll}
  b_{12}=0\,,
& b_{14}=    +\left(\frac{m_sm_b}{M_W^2}\right)^2\frac{m_s}{M_W}\,,
 & \hbox{at } E=m_s
\\b_{13}\sim -\left(\frac{m_sm_b}{M_W^2}\right)^2
   \frac{9}{4}\frac{m_s^2}{m_bM_W},
& b_{14}\sim +\left(\frac{m_sm_b}{M_W^2}\right)^2
   \frac{3}{2}\frac{m_s^2}{m_bM_W}\,,
 & \hbox{at } E=m_b.
\end{array}
\label{btos}
\ee

The total asymmetry, eq. (\ref{CPas2}), is clearly dominated by the
$s_R\rightarrow b_L$ contribution, with a maximum value
$\Delta_{CP}^{T=0}(\omega\sim m_b)=5\,10^{-26}$ around
the $m_b$ threshold.

Consider now charge $-2/3$ quarks. Strictly speaking, our computation is not a
good approximation for this sector as in eq.(\ref{lws2}) we have neglected
factors of $m_i^2/M_W^2$, $m_f^2/M_W^2$, etc., which is not legitimate for the
top quark. Nevertheless the type of GIM cancellations, which we
justified in terms of general arguments in Sect. \ref{secloop}, should still
hold. In particular, the analytical behaviour for the internal mass dependence
($d,s,b$ in this case), given in eq. (\ref{sap}), are specially accurate
because all internal masses are really small in front of $M_W$. The exact
numerical result for these quantities is
\be
\begin{array}{ll}
 \sum S_{1,2}(M_l,M_{l'})=\,2.37\,10^{-10}
&\sum S_{1,3}(M_l,M_{l'})=-1.12\,10^{-10}\\
 \sum S_{1,4}(M_l,M_{l'})=\,3.33\,10^{-10}
&\sum S_{2,3}(M_l,M_{l'})=-7.37\,10^{-12}\\
 \sum S_{2,4}(M_l,M_{l'})=\,4.9\,10^{-11}
&\sum S_{3,4}(M_l,M_{l'})=3.45 \,10^{-11},
\end{array}
\ee
where the sums now run on $l,l'= d,s,b$ under the usual cyclicity
constraint $l'=l+1 \pmod3$.

In this toy computation, the ``up" type quarks contribution is expected to be
be about the same as the ``down" type one: while the external mass dependence
gives a bigger factor, the internal GIM cancellation produces the opposite
effect. Applying for the sake of the argument our analytical formulae to this
sector, gives an external mass dependence at the dominant threshold $\sim
m_q^5/M_W^5$, while the internal GIM cancellation goes as $\sim
m_q^6/M_W^6$. The contributions of both types of quarks should not differ
much. Using our formulae at face value results in a dominant $c_R\rightarrow
t_L$ contribution of order $\Delta^{T=0}_{CP}(\omega\sim m_t)\sim
\,2.5\,10^{-25}$.  In the present case
$b_{12}$ is not zero, which reinforces the effect.

As previously discussed, in a complete computation, an external suppression
factor $\sim (\frac{m_i m_f}{M_W^2})^2$ could disappear, leading to a large
enhancement for certain flavours (up to 13 orders of magnitude for the $d-s$
pair, still insufficient for the observed asymmetry), although this may be
very optimistic as the pattern of unitarity triangles must persist.

\seCtion{Conclusions}

 Realistic models of baryogenesis imply the inclusion of finite temperature
effects. We consider scenarios with a first order phase transition. It is
interesting to disentangle whether temperature effects, other than the
existence of the transition itself, are responsible for possible modifications
of standard CP arguments. We thus revert in a first step to the analysis of
particle propagation at $T=0$.

The essential non perturbative element is the wall separating the two phases
of spontaneous symmetry breaking. The propagation of any particle of the SM
spectrum should be exactly solved in its presence. And this we have done for a
free fermion and a thin wall, leading to a new Feynman
propagator which replaces and generalizes the usual one. It contains both
massless and massive poles, and new CP-even phases, present both for on-shell
and off-shell fermions. They stem from the tree-level reflection of fermions
on the bubble wall. This propagator can be useful in physical process other
than baryogenesis, where a first order phase transition is relevant.

When electroweak loops are considered, a CP-asymmetry is found in the one-loop
reflection properties of quarks and antiquarks. The one-loop self-energies
cannot be completely renormalized away for on-shell fermions in the presence
of a wall, as the latter acts as a source of momenta. The asymmetry is then
found at order $\alpha_W$ in amplitude and requires the interference of the
mentioned CP-even phases with the CP-odd CKM ones. The desired effect is thus
present without any consideration of temperature effects in the quark
propagation.

 A complete one-loop SM calculation is beyond the scope of the present
paper. We have instead performed a toy computation in which only the broken
phase is considered inside the loops, in a Lorentz covariant way. The result
is many orders of magnitude below what observation requires, and it can be
expressed in terms of two unitarity triangles. They show the internal and
external GIM cancellations and the interplay between CP-even and CP-odd
phases. This pattern of triangles should survive in a complete computation,
which we leave for later publication. We argue that suppression factors in
external masses could be absent then, due to the breaking of Lorentz
invariance in the loops analysis, although the quantitative enhancement is
insufficient.

With the expertise acquired in this paper, we face in ref.\cite{nousT} the
 physical finite temperature case, where we discuss
 the common building blocks and new features of the scenario.

\section*{Acknowledgments}

 We acknowledge Luis Alvarez-Gaum\'e, Alvaro De R\'ujula, 
Jean-Marie Fr\`ere, Pilar Hern\'andez and Jean-Pierre Leroy  for many inspiring discussions. We are specially indebted to Andy
Cohen, who took part in the early discussions. M.B. Gavela is indebted to ITP
(Santa Barbara)
for hospitality in the final period of this work, where her research was
supported in part
by the National Science Foundation under Grant No. PHY89-04035.

\appendix
\seCtion{Feynman boundary conditions for the fermion propagator}
\label{app-proof}

 We prove here that the propagator derived in  eqs. (\ref{prol}), (\ref{pror})
and (\ref{propsym})
is a unique solution of the inhomogeneous Dirac equation (\ref{green2}).
 In a first step,  the equivalence between
eqs. (\ref{prol}) and (\ref{pror}) will be established. We then prove that the
two `$i \epsilon$''
conventions are indeed equivalent, as stated in section (\ref{secprop}). Next
we show,
 in the ``space $i\epsilon$'' convention, that the propagator satisfies the
 inhomogeneous Dirac equation. Finally, in the (Feynman) ``time $i\epsilon$''
convention, we demonstrate
 that the Feynman boundary
conditions are indeed satisfied.

\subsection{Equivalence between eqs. (4.16) and (4.17).}
\label{sec-equi}

Let us consider the difference between the right hand sides of
 eq.  (\ref{pror}) and eq.
(\ref{prol}).
 The difference between the first lines of these equations is

\be
\frac{1}{q^f_z-q^i_z+i\epsilon} \left (\frac{1}{\dsl q^i} - \frac{1}{\dsl
q^f}\right)
-\frac{1}{q^f_z-q^i_z-i\epsilon} \left(\frac{1}{\dsl q^i-m}-\frac{1}{\dsl
q^f-m}\right),
\label{linun}\ee
where  the Feynman convention is understood in all  all $1/\dsl q$',
 $1/(\dsl q-m)$'s.

The difference between the second and third lines of the same equations is

\be
-\frac{1}{\dsl q^f} s \alpha_z s \gamma_0 \frac{1}{\dsl q^i}
+\frac{1}{\dsl q^f-m} s \alpha_z s \gamma_0 \frac{1}{\dsl q^i-m}.
\label{lindt}\ee

 $q^f$ and $q^i$ differ only in their $z$ component, implying that
  $\dsl q^f -\dsl q^i= -\gamma_z (q^f_z-q^i)$. Using this equality, together
with
 $1/{\dsl q^i} - 1/{\dsl q^f}=1/{\dsl q^f}
(\dsl q^f -\dsl q^i)1/{\dsl q^i}$, it is straightforw to show  that the sum of
eqs.
(\ref{linun}) and (\ref{lindt}) vanishes, whence the equality of r.h.s. of
(\ref{prol})
and of (\ref{pror}).

\subsection{Equivalence between the two ``$i\epsilon$'' conventions.}
\label{sec-eps}

The usual``time $i\epsilon$'' simply amounts to $q_0^2 \ra q_0^2 + i\epsilon$,
while the ``space $i\epsilon$'' convention has been defined in Sec.
\ref{secprop}.  The difference between the two
``$i\epsilon$'' conventions has been exhibited in eq. (\ref{ie}).

Consider the equality

$$ -\frac m {\dsl q^f (\dsl q^f-m)} = -\frac 1 {\dsl q^f-m}+\frac 1{\dsl q^f}.
$$

 Aplying it to eq. (\ref{pror}), we can separate the terms containing  $1/{\dsl
q^f}$ and those containing $1/{\dsl q^f-m}$. Let us dismiss momentarily the
latter, which will be discussed later.

It is possible to prove that the substitution of all $1/{\dsl q^f}$ factors in
eq. (\ref{pror} by  $2 i \pi \dsl q^f \theta(q_z^f)\delta(q_0^2-(q_z^f)^2)$
gives a vanishing result. Indeed, from

\be \dsl q^f \theta(q_z^f)\delta(q_0^2-(q_z^f)^2) = s\vert q_0\vert \gamma^0
(1-s\alpha_z) \theta(q_z^f)\delta(q_0^2-(q_z^f)^2),\label{psix}\ee
the third line in eq. (\ref{pror}) gives then a vanishing coefficient, since
$(1-s\alpha_z)(1+s\alpha_z)=0$. Applying now in the second line the equality

\be (1-s\alpha_z)\left[1-\frac{m s \gamma_0}{E+p_z'}\right]\frac{1-s\alpha_z}2
=
1-s\alpha_z, \label{psept}\ee
it follows that

\be (1-s\alpha_z)s \gamma_0 \frac{1}{\dsl q^i}=s(1-s\alpha_z)\frac
{q^f_z+q^i_z}
{(q^f_z)^2-(q^i_z)^2+i\epsilon}=s(1-s\alpha_z) \frac 1 {q^f_z-q^i_z+i\epsilon},
\label{phuit}\ee
where we have used $sq_0=q^f_z$ (since $sq_0>0$ by definition of $s$, and
$q^f_z>0$ due to the
$\theta(q_z^f)$ factor). Finally, using once more  eq. (\ref {psix}), the
second
line gives

\be  \dsl q^f \theta(q_z^f)\delta(q_0^2-(q_z^f)^2) \frac 1
{q^f_z-q^i_z+i\epsilon},
\label{seclin}\ee.
which obviously cancels the contribution from the first line.

Let us now turn to the $1/(\dsl q^f -m)$ term. The equivalent of eq. (\ref{ie})
is

\be
\frac {\dsl q^f+m}{q_0^2-(q_z^f+i \epsilon)^2-m^2}=
\frac {\dsl q^f+m}{q_0^2-(q_z^f)^2+i \epsilon-m^2}
+ 2 i \pi (\dsl q^f +m)\theta(-q_z^f)\delta(q_0^2-(q_z^f)^2-m^2).\label{ie}
\ee

The proof proceeds in a way similar to the preceding one but slightly more
involved.
 It can be shown that

 \be (\dsl q^f +m)\theta(-q_z^f)\delta(q_0^2-(q_z^f)^2-m^2)=
s\vert q_0\vert \gamma^0
\left(1 +\frac{p_z^f \alpha_z + m\gamma^0}{s\vert
q_0\vert}\right),\label{projj}\ee
where $p_z^f=\sqrt{q_0^2-m^2}$. The last bracket in (\ref{projj}) is twice a
projector
that plays the same role as $1-s\alpha_z$ in the preceding derivation.

Using
\be \left(1 +\frac{p_z^f \alpha_z + m\gamma^0}{s\vert
q_0\vert}\right) \left[1-\frac{m s \gamma_0}{E+p_z'}\right]
 \frac{1- s \alpha_z}{2}=0,\ee
it follows  that the contribution from the second line of (\ref{pror})
vanishes.
{}From

\be \left(1 +\frac{p_z^f \alpha_z + m\gamma^0}{s\vert
q_0\vert}\right) \frac{1+ s \alpha_z}{2}\left[1+\frac{m s
\gamma_0}{E+p_z'}\right]
 =\left(1 +\frac{p_z^f \alpha_z + m\gamma^0}{s\vert
q_0\vert}\right)\ee
it is possible to check  that the third line cancels the contribution from
the first one, which ends the proof for the  for $1/(\dsl q^f -m)$ term.

The $1/(\dsl q^i -m)$ and $1/\dsl q^i$ may be treated in a similar way leading
to the
anounced conclusion: the quark propagator in eq. (\ref{pror}) has the same
value
whichever ``$i \epsilon$'' convention is used. From the equality derived in the
preceding subsection, this applies also to eqs. (\ref{prol}) and
(\ref{propsym}).

\subsection{The inhomogeneous Dirac equation.}
\label{sec-dir}

 We prove here that the propagator verifies the inhomogeneous Dirac equation.
Let us consider the equation (\ref{green2}). Multiplying both sides by
$i\gamma_0$, the
Dirac operator becomes

\be \theta(-\xi_z')i\dsl \partial + \theta(\xi_z')(i\dsl \partial
-m)\label{dirop}.\ee
 We use the ``space $i\epsilon$'' convention, so that  $1/\dsl q^f$ has a pole
in
the $q_z^f$ complex plane below the real axis. Its Fourier transform has then a
pole above the real axis,

\be\int dq^f_z e^{i q^f_z \xi_z'} \frac 1 {\dsl q^f} \propto \theta(-\xi_z')
[i\dsl \partial]^{-1}.\ee
Similarly,  $1/(\dsl q^f-m)$ has a pole above the real axis, leading to

\be\int dq^f_z e^{i q^f_z \xi_z'} \frac 1 {\dsl q^f-m} \propto \theta(\xi_z')
[i\dsl \partial-m]^{-1}.\ee

As a consequence, the operator  (\ref{dirop}) applied to $m/\dsl q^f (\dsl
q^f-m)$ gives zero. It follows that the second and third lines in
eq. (\ref{pror}) vanish under the action of the Dirac operator. Its action on
the first line
gives

\be -\frac{1}{q^f_z-q^i_z+i\epsilon}  +
 \frac{1}{q^f_z-q^i_z-i\epsilon} = 2 i\pi \delta(q^f_z-q^i_z). \ee

This ends the proof. The same argument holds when the Dirac equation is applied
on the right
hand side of the propagator. In this case, it is more convenient to perform the
analysis on eq. (\ref{prol}), sticking again to the ``space $i\epsilon$''
convention.

\subsection{Feynman boundary conditions.}
\label{sec-feyn}

In the ``time $i\epsilon$'' convention, it is trivial to proof that the
propagator verifies the Feynman boundary conditions.  Indeed, the latter state
that the positive frequencies are ``retarded'' while the
negative ones are ``advanced''. By definition, the ``time $i\epsilon$''
prepscription achieves it, as it amounts to the replacement  $q_0^2\ra q_0^2
+i \epsilon$, which  leads to poles in the $q_0$ complex plane below (above)
the real axis
when its real part is positive (negative).

\end{document}

%% file: psbox.tex
\def\temp{1.35}%
\let\tempp=\relax
\expandafter\ifx\csname psboxversion\endcsname\relax
  \message{PSBOX(\temp)}%
\else
    \ifdim\temp cm>\psboxversion cm
      \message{PSBOX(\temp)}%
    \else
      \message{PSBOX(\psboxversion) is already loaded: I won't load
        PSBOX(\temp)!}%
      \let\temp=\psboxversion
      \let\tempp= 
    \fi
\fi
\tempp
\message{by Jean Orloff: loading ...}
\let\psboxversion=\temp
\catcode`\@=11
%
%
\def\psfortextures{
\def\PSspeci@l##1##2{%
\special{illustration ##1\space scaled ##2}%
}}%
\def\psfordvitops{
\def\PSspeci@l##1##2{%
\special{dvitops: import ##1\space \the\drawingwd \the\drawinght}%
}}%
\def\psfordvips{
\def\PSspeci@l##1##2{%
\d@my=0.1bp \d@mx=\drawingwd \divide\d@mx by\d@my
\includegraphics{##1\space}}}%
\def\psforoztex{
\def\PSspeci@l##1##2{%
\special{##1 \space
      ##2 1000 div dup scale
      \number-\psllx\space\space \number-\pslly\space\space translate
}}}%
\def\psfordvitps{
\def\dvitpsLiter@ldim##1{\dimen0=##1\relax
\special{dvitps: Literal "\number\dimen0\space"}}%
\def\PSspeci@l##1##2{%
\at(0bp;\drawinght){%
\special{dvitps: Include0 "psfig.psr"}
\dvitpsLiter@ldim{\drawingwd}%
\dvitpsLiter@ldim{\drawinght}%
\dvitpsLiter@ldim{\psllx bp}%
\dvitpsLiter@ldim{\pslly bp}%
\dvitpsLiter@ldim{\psurx bp}%
\dvitpsLiter@ldim{\psury bp}%
\special{dvitps: Literal "startTexFig"}%
\special{dvitps: Include1 "##1"}%
\special{dvitps: Literal "endTexFig"}%
}}}%
\def\psfordvialw{
\def\PSspeci@l##1##2{
\special{language "PostScript",
position = "bottom left",
literal "  \psllx\space \pslly\space translate
  ##2 1000 div dup scale
  -\psllx\space -\pslly\space translate",
include "##1"}
}}%
\def\psforptips{
\def\PSspeci@l##1##2{{
\d@mx=\psurx bp
\advance \d@mx by -\psllx bp
\divide \d@mx by 1000\multiply\d@mx by \xscale
\incm{\d@mx}
\let\tmpx\dimincm
\d@my=\psury bp
\advance \d@my by -\pslly bp
\divide \d@my by 1000\multiply\d@my by \xscale
\incm{\d@my}
\let\tmpy\dimincm
\d@mx=-\psllx bp
\divide \d@mx by 1000\multiply\d@mx by \xscale
\d@my=-\pslly bp
\divide \d@my by 1000\multiply\d@my by \xscale
\at(\d@mx;\d@my){\special{ps:##1 x=\tmpx cm, y=\tmpy cm}}
}}}%
\def\psonlyboxes{
\def\PSspeci@l##1##2{%
\at(0cm;0cm){\boxit{\vbox to\drawinght
  {\vss\hbox to\drawingwd{\at(0cm;0cm){\hbox{({\tt##1})}}\hss}}}}
}}%
\def\psloc@lerr#1{%
\let\savedPSspeci@l=\PSspeci@l%
\def\PSspeci@l##1##2{%
\at(0cm;0cm){\boxit{\vbox to\drawinght
  {\vss\hbox to\drawingwd{\at(0cm;0cm){\hbox{({\tt##1}) #1}}\hss}}}}
\let\PSspeci@l=\savedPSspeci@l
}}%
%
%
\newread\pst@mpin
\newdimen\drawinght\newdimen\drawingwd
\newdimen\psxoffset\newdimen\psyoffset
\newbox\drawingBox
\newcount\xscale \newcount\yscale \newdimen\pscm\pscm=1cm
\newdimen\d@mx \newdimen\d@my
\newdimen\pswdincr \newdimen\pshtincr
\let\ps@nnotation=\relax
{\catcode`\|=0 |catcode`|\=12 |catcode`|
|catcode`#=12 |catcode`*=14
|xdef|backslashother{\}*
|xdef|percentother{
|xdef|tildeother{~}*
|xdef|sharpother{#}*
}%
\def\R@moveMeaningHeader#1:->{}%
\def\uncatcode#1{%
\edef#1{\expandafter\R@moveMeaningHeader\meaning#1}}%
\def\execute#1{#1}
\def\psm@keother#1{\catcode`#112\relax}
\def\executeinspecs#1{%
\execute{\begingroup\let\do\psm@keother\dospecials\catcode`\^^M=9#1\endgroup}}%
\def\@mpty{}%
\def\matchexpin#1#2{
  \fi%
  \edef\tmpb{{#2}}%
  \expandafter\makem@tchtmp\tmpb%
  \edef\tmpa{#1}\edef\tmpb{#2}%
  \expandafter\expandafter\expandafter\m@tchtmp\expandafter\tmpa\tmpb\endm@tch%
  \if\match%
}%
\def\matchin#1#2{%
  \fi%
  \makem@tchtmp{#2}%
  \m@tchtmp#1#2\endm@tch%
  \if\match%
}%
\def\makem@tchtmp#1{\def\m@tchtmp##1#1##2\endm@tch{%
  \def\tmpa{##1}\def\tmpb{##2}\let\m@tchtmp=\relax%
  \ifx\tmpb\@mpty\def\match{YN}%
  \else\def\match{YY}\fi%
}}%
\def\incm#1{{\psxoffset=1cm\d@my=#1
 \d@mx=\d@my
  \divide\d@mx by \psxoffset
  \xdef\dimincm{\number\d@mx.}
  \advance\d@my by -\number\d@mx cm
  \multiply\d@my by 100
 \d@mx=\d@my
  \divide\d@mx by \psxoffset
  \edef\dimincm{\dimincm\number\d@mx}
  \advance\d@my by -\number\d@mx cm
  \multiply\d@my by 100
 \d@mx=\d@my
  \divide\d@mx by \psxoffset
  \xdef\dimincm{\dimincm\number\d@mx}
}}%
%
\newif\ifNotB@undingBox
\newhelp\PShelp{Proceed: you'll have a 5cm square blank box instead of
your graphics.}%
\def\s@tsize#1 #2 #3 #4\@ndsize{
  \def\psllx{#1}\def\pslly{#2}%
  \def\psurx{#3}\def\psury{#4}
  \ifx\psurx\@mpty\NotB@undingBoxtrue
  \else
    \drawinght=#4bp\advance\drawinght by-#2bp
    \drawingwd=#3bp\advance\drawingwd by-#1bp
  \fi
  }%
\def\sc@nBBline#1:#2\@ndBBline{\edef\p@rameter{#1}\edef\v@lue{#2}}%
\def\g@bblefirstblank#1#2:{\ifx#1 \else#1\fi#2}%
{\catcode`\%=12
\xdef\B@undingBox{
\def\ReadPSize#1{
 \readfilename#1\relax
 \let\PSfilename=\lastreadfilename
 \openin\pst@mpin=#1\relax
 \ifeof\pst@mpin \errhelp=\PShelp
   \errmessage{I haven't found your postscript file (\PSfilename)}%
   \psloc@lerr{was not found}%
   \s@tsize 0 0 142 142\@ndsize
   \closein\pst@mpin
 \else
   \if\matchexpin{\GlobalInputList}{, \lastreadfilename}%
   \else\xdef\GlobalInputList{\GlobalInputList, \lastreadfilename}%
     \immediate\write\psbj@inaux{\lastreadfilename,}%
   \fi%
   \loop
     \executeinspecs{\catcode`\ =10\global\read\pst@mpin to\n@xtline}%
     \ifeof\pst@mpin
       \errhelp=\PShelp
       \errmessage{(\PSfilename) is not an Encapsulated PostScript File:
           I could not find any \B@undingBox: line.}%
       \edef\v@lue{0 0 142 142:}%
       \psloc@lerr{is not an EPSFile}%
       \NotB@undingBoxfalse
     \else
       \expandafter\sc@nBBline\n@xtline:\@ndBBline
       \ifx\p@rameter\B@undingBox\NotB@undingBoxfalse
         \edef\t@mp{%
           \expandafter\g@bblefirstblank\v@lue\space\space\space}%
         \expandafter\s@tsize\t@mp\@ndsize
       \else\NotB@undingBoxtrue
       \fi
     \fi
   \ifNotB@undingBox\repeat
   \closein\pst@mpin
 \fi
\message{#1}%
}%
%
%
\def\psboxto(#1;#2)#3{\vbox{%
   \ReadPSize{#3}%
   \advance\pswdincr by \drawingwd
   \advance\pshtincr by \drawinght
   \divide\pswdincr by 1000
   \divide\pshtincr by 1000
   \d@mx=#1
   \ifdim\d@mx=0pt\xscale=1000
         \else \xscale=\d@mx \divide \xscale by \pswdincr\fi
   \d@my=#2
   \ifdim\d@my=0pt\yscale=1000
         \else \yscale=\d@my \divide \yscale by \pshtincr\fi
   \ifnum\yscale=1000
         \else\ifnum\xscale=1000\xscale=\yscale
                    \else\ifnum\yscale<\xscale\xscale=\yscale\fi
              \fi
   \fi
   \divide\drawingwd by1000 \multiply\drawingwd by\xscale
   \divide\drawinght by1000 \multiply\drawinght by\xscale
   \divide\psxoffset by1000 \multiply\psxoffset by\xscale
   \divide\psyoffset by1000 \multiply\psyoffset by\xscale
   \global\divide\pscm by 1000
   \global\multiply\pscm by\xscale
   \multiply\pswdincr by\xscale \multiply\pshtincr by\xscale
   \ifdim\d@mx=0pt\d@mx=\pswdincr\fi
   \ifdim\d@my=0pt\d@my=\pshtincr\fi
   \message{scaled \the\xscale}%
 \hbox to\d@mx{\hss\vbox to\d@my{\vss
   \global\setbox\drawingBox=\hbox to 0pt{\kern\psxoffset\vbox to 0pt{%
      \kern-\psyoffset
      \PSspeci@l{\PSfilename}{\the\xscale}%
      \vss}\hss\ps@nnotation}%
   \global\wd\drawingBox=\the\pswdincr
   \global\ht\drawingBox=\the\pshtincr
   \global\drawingwd=\pswdincr
   \global\drawinght=\pshtincr
   \baselineskip=0pt
   \copy\drawingBox
 \vss}\hss}%
  \global\psxoffset=0pt
  \global\psyoffset=0pt
  \global\pswdincr=0pt
  \global\pshtincr=0pt 
  \global\pscm=1cm 
}}%
%
%
\def\psboxscaled#1#2{\vbox{%
  \ReadPSize{#2}%
  \xscale=#1
  \message{scaled \the\xscale}%
  \divide\pswdincr by 1000 \multiply\pswdincr by \xscale
  \divide\pshtincr by 1000 \multiply\pshtincr by \xscale
  \divide\psxoffset by1000 \multiply\psxoffset by\xscale
  \divide\psyoffset by1000 \multiply\psyoffset by\xscale
  \divide\drawingwd by1000 \multiply\drawingwd by\xscale
  \divide\drawinght by1000 \multiply\drawinght by\xscale
  \global\divide\pscm by 1000
  \global\multiply\pscm by\xscale
  \global\setbox\drawingBox=\hbox to 0pt{\kern\psxoffset\vbox to 0pt{%
     \kern-\psyoffset
     \PSspeci@l{\PSfilename}{\the\xscale}%
     \vss}\hss\ps@nnotation}%
  \advance\pswdincr by \drawingwd
  \advance\pshtincr by \drawinght
  \global\wd\drawingBox=\the\pswdincr
  \global\ht\drawingBox=\the\pshtincr
  \global\drawingwd=\pswdincr
  \global\drawinght=\pshtincr
  \baselineskip=0pt
  \copy\drawingBox
  \global\psxoffset=0pt
  \global\psyoffset=0pt
  \global\pswdincr=0pt
  \global\pshtincr=0pt 
  \global\pscm=1cm
}}%
%
\def\psbox#1{\psboxscaled{1000}{#1}}%
\newif\ifn@teof\n@teoftrue
\newif\ifc@ntrolline
\newif\ifmatch
\newread\j@insplitin
\newwrite\j@insplitout
\newwrite\psbj@inaux
\immediate\openout\psbj@inaux=psbjoin.aux
\immediate\write\psbj@inaux{\string\joinfiles}%
\immediate\write\psbj@inaux{\jobname,}%
%
%
\def\toother#1{\ifcat\relax#1\else\expandafter%
  \toother@ux\meaning#1\endtoother@ux\fi}%
\def\toother@ux#1 #2#3\endtoother@ux{\def\tmp{#3}%
  \ifx\tmp\@mpty\def\tmp{#2}\let\next=\relax%
  \else\def\next{\toother@ux#2#3\endtoother@ux}\fi%
\next}%
%
%
\let\readfilenamehook=\relax
\def\re@d{\expandafter\re@daux}
\def\re@daux{\futurelet\nextchar\stopre@dtest}%
\def\re@dnext{\xdef\lastreadfilename{\lastreadfilename\nextchar}%
  \afterassignment\re@d\let\nextchar}%
\def\stopre@d{\egroup\readfilenamehook}%
\def\stopre@dtest{%
  \ifcat\nextchar\relax\let\nextread\stopre@d
  \else
    \ifcat\nextchar\space\def\nextread{%
      \afterassignment\stopre@d\chardef\nextchar=`}%
    \else\let\nextread=\re@dnext
      \toother\nextchar
      \edef\nextchar{\tmp}%
    \fi
  \fi\nextread}%
\def\readfilename{\bgroup%
  \let\\=\backslashother \let\%=\percentother \let\~=\tildeother
  \let\#=\sharpother \xdef\lastreadfilename{}%
  \re@d}%
%
%
\xdef\GlobalInputList{\jobname}%
\def\psnewinput{%
  \def\readfilenamehook{
    \if\matchexpin{\GlobalInputList}{, \lastreadfilename}%
    \else\xdef\GlobalInputList{\GlobalInputList, \lastreadfilename}%
      \immediate\write\psbj@inaux{\lastreadfilename,}%
    \fi%
    \let\readfilenamehook=\relax%
    \ps@ldinput\lastreadfilename\relax%
  }\readfilename%
}%
\expandafter\ifx\csname @@input\endcsname\relax    
  \immediate\let\ps@ldinput=\input\def\input{\psnewinput}%
\else
  \immediate\let\ps@ldinput=\@@input
  \def\@@input{\psnewinput}%
\fi%
\def\nowarnopenout{%
 \def\warnopenout##1##2{%
   \readfilename##2\relax
   \message{\lastreadfilename}%
   \immediate\openout##1=\lastreadfilename\relax}}%
\def\warnopenout#1#2{%
 \readfilename#2\relax
 \def\t@mp{TrashMe,psbjoin.aux,psbjoint.tex,}\uncatcode\t@mp
 \if\matchexpin{\t@mp}{\lastreadfilename,}%
 \else
   \immediate\openin\pst@mpin=\lastreadfilename\relax
   \ifeof\pst@mpin
     \else
     \edef\tmp{{If the content of this file is precious to you, this
is your last chance to abort (ie press x or e) and rename it before
retexing (\jobname). If you're sure there's no file
(\lastreadfilename) in the directory of (\jobname), then go on: I'm
simply worried because you have another (\lastreadfilename) in some
directory I'm looking in for inputs...}}%
     \errhelp=\tmp
     \errmessage{I may be about to replace your file named \lastreadfilename}%
   \fi
   \immediate\closein\pst@mpin
 \fi
 \message{\lastreadfilename}%
 \immediate\openout#1=\lastreadfilename\relax}%
{\catcode`\%=12\catcode`\*=14
\gdef\splitfile#1{*
 \readfilename#1\relax
 \immediate\openin\j@insplitin=\lastreadfilename\relax
 \ifeof\j@insplitin
   \message{! I couldn't find and split \lastreadfilename!}*
 \else
   \immediate\openout\j@insplitout=TrashMe
   \message{< Splitting \lastreadfilename\space into}*
   \loop
     \ifeof\j@insplitin
       \immediate\closein\j@insplitin\n@teoffalse
     \else
       \n@teoftrue
       \executeinspecs{\global\read\j@insplitin to\spl@tinline\expandafter
         \ch@ckbeginnewfile\spl@tinline
       \ifc@ntrolline
       \else
         \toks0=\expandafter{\spl@tinline}*
         \immediate\write\j@insplitout{\the\toks0}*
       \fi
     \fi
   \ifn@teof\repeat
   \immediate\closeout\j@insplitout
 \fi\message{>}*
}*
\gdef\ch@ckbeginnewfile#1
 \def\t@mp{#1}*
 \ifx\@mpty\t@mp
   \def\t@mp{#3}*
   \ifx\@mpty\t@mp
     \global\c@ntrollinefalse
   \else
     \immediate\closeout\j@insplitout
     \warnopenout\j@insplitout{#2}*
     \global\c@ntrollinetrue
   \fi
 \else
   \global\c@ntrollinefalse
 \fi}*
\gdef\joinfiles#1\into#2{*
 \message{< Joining following files into}*
 \warnopenout\j@insplitout{#2}*
 \message{:}*
 {*
 \edef\w@##1{\immediate\write\j@insplitout{##1}}*
\w@{
\w@{
\w@{
\w@{
\w@{
\w@{
\w@{
\w@{
\w@{
\w@{
\w@{\string\input\space psbox.tex}*
\w@{\string\splitfile{\string\jobname}}*
\w@{\string\let\string\autojoin=\string\relax}*
}*
 \expandafter\tre@tfilelist#1, \endtre@t
 \immediate\closeout\j@insplitout
 \message{>}*
}*
\gdef\tre@tfilelist#1, #2\endtre@t{*
 \readfilename#1\relax
 \ifx\@mpty\lastreadfilename
 \else
   \immediate\openin\j@insplitin=\lastreadfilename\relax
   \ifeof\j@insplitin
     \errmessage{I couldn't find file \lastreadfilename}*
   \else
     \message{\lastreadfilename}*
     \immediate\write\j@insplitout{
     \executeinspecs{\global\read\j@insplitin to\oldj@ininline}*
     \loop
       \ifeof\j@insplitin\immediate\closein\j@insplitin\n@teoffalse
       \else\n@teoftrue
         \executeinspecs{\global\read\j@insplitin to\j@ininline}*
         \toks0=\expandafter{\oldj@ininline}*
         \let\oldj@ininline=\j@ininline
         \immediate\write\j@insplitout{\the\toks0}*
       \fi
     \ifn@teof
     \repeat
   \immediate\closein\j@insplitin
   \fi
   \tre@tfilelist#2, \endtre@t
 \fi}*
}%
\def\autojoin{%
 \immediate\write\psbj@inaux{\string\into{psbjoint.tex}}%
 \immediate\closeout\psbj@inaux
 \expandafter\joinfiles\GlobalInputList\into{psbjoint.tex}%
}%
%
%
%
\def\centinsert#1{\midinsert\line{\hss#1\hss}\endinsert}%
\def\psannotate#1#2{\vbox{%
  \def\ps@nnotation{#2\global\let\ps@nnotation=\relax}#1}}%
\def\pscaption#1#2{\vbox{%
   \setbox\drawingBox=#1
   \copy\drawingBox
   \vskip\baselineskip
   \vbox{\hsize=\wd\drawingBox\setbox0=\hbox{#2}%
     \ifdim\wd0>\hsize
       \noindent\unhbox0\tolerance=5000
    \else\centerline{\box0}%
    \fi
}}}%
%
\def\at(#1;#2)#3{\setbox0=\hbox{#3}\ht0=0pt\dp0=0pt
  \rlap{\kern#1\vbox to0pt{\kern-#2\box0\vss}}}%
%
\newdimen\gridht \newdimen\gridwd
\def\gridfill(#1;#2){%
  \setbox0=\hbox to 1\pscm
  {\vrule height1\pscm width.4pt\leaders\hrule\hfill}%
  \gridht=#1
  \divide\gridht by \ht0
  \multiply\gridht by \ht0
  \gridwd=#2
  \divide\gridwd by \wd0
  \multiply\gridwd by \wd0
  \advance \gridwd by \wd0
  \vbox to \gridht{\leaders\hbox to\gridwd{\leaders\box0\hfill}\vfill}}%
%
\def\fillinggrid{\at(0cm;0cm){\vbox{%
  \gridfill(\drawinght;\drawingwd)}}}%
%
%
\def\textleftof#1:{%
  \setbox1=#1
  \setbox0=\vbox\bgroup
    \advance\hsize by -\wd1 \advance\hsize by -2em}%
\def\textrightof#1:{%
  \setbox0=#1
  \setbox1=\vbox\bgroup
    \advance\hsize by -\wd0 \advance\hsize by -2em}%
\def\endtext{%
  \egroup
  \hbox to \hsize{\valign{\vfil##\vfil\cr%
\box0\cr%
\noalign{\hss}\box1\cr}}}%
%
\def\frameit#1#2#3{\hbox{\vrule width#1\vbox{%
  \hrule height#1\vskip#2\hbox{\hskip#2\vbox{#3}\hskip#2}%
        \vskip#2\hrule height#1}\vrule width#1}}%
\def\boxit#1{\frameit{0.4pt}{0pt}{#1}}%
\catcode`\@=12 
%
\psfordvips   